\documentclass[
 reprint,
 showpacs, 
 amsmath, amssymb,
 aps,
 prl,
 floatfix,
 superscriptaddress
]{revtex4-2}

\usepackage{amsmath}
\usepackage{graphicx}
\usepackage{dcolumn}
\usepackage{bm}
\usepackage{bbm}
\usepackage{comment}

\usepackage[utf8]{inputenc}
\usepackage[T1]{fontenc}
\usepackage{textgreek}
\usepackage{upgreek}
\usepackage{float}
\usepackage{hyperref}

\usepackage{newtxtext}
\usepackage{newtxmath}

\usepackage{color}
\usepackage[normalem]{ulem}
\usepackage{xcolor}
\definecolor{darkgreen}{RGB}{0 100 0}

\usepackage{natbib}

\usepackage{amsmath,amssymb}
\usepackage{mathrsfs}
\usepackage{physics}
\newcommand{\beginsupplement}{%
        \setcounter{table}{0}
        \renewcommand{\thetable}{S\arabic{table}}%
        \setcounter{figure}{0}
        \renewcommand{\thefigure}{S\arabic{figure}}%
				\renewcommand{\theequation}{S.\arabic{equation}}
     }
		
\begin{document}

	\preprint{}

\title[Ratchet effect in JJA]{Spontaneous supercurrents and vortex depinning in two-dimensional arrays of $\varphi_0$-junctions}

\author{S.~Reinhardt}
\affiliation{Institut f\"ur Experimentelle und Angewandte Physik, University of Regensburg, 93040 Regensburg, Germany}
\affiliation{School of Applied and Engineering Physics, Cornell University, Ithaca, NY, 14853, USA}
\author{A.~G.~Penner}
\affiliation{Dahlem Center for Complex Quantum Systems and Fachbereich Physik, Freie Universität Berlin, 14195 Berlin, Germany}

\author{J.~Berger}
\author{C.~Baumgartner}

\affiliation{Institut f\"ur Experimentelle und Angewandte Physik, University of Regensburg, 93040 Regensburg, Germany}

\author{S.~Gronin}
\author{G.~C.~Gardner}
\affiliation{Birck Nanotechnology Center, Purdue University, West Lafayette, Indiana 47907 USA}
\author{T.~Lindemann}
\affiliation{Birck Nanotechnology Center, Purdue University, West Lafayette, Indiana 47907 USA}
\affiliation{Department of Physics and Astronomy, Purdue University, West Lafayette, Indiana 47907 USA}

\author{M.~J.~Manfra}
\affiliation{Birck Nanotechnology Center, Purdue University, West Lafayette, Indiana 47907 USA}
\affiliation{Department of Physics and Astronomy, Purdue University, West Lafayette, Indiana 47907 USA}
\affiliation{School of Materials Engineering, Purdue University, West Lafayette, Indiana 47907 USA}
\affiliation{Elmore Family School of Electrical and Computer Engineering, Purdue University, West Lafayette, Indiana 47907 USA}

\author{L.~I.~Glazman}
\affiliation{Department of Physics, Yale University, New Haven, Connecticut 06520, USA}

\author{F.~von Oppen}
\affiliation{Dahlem Center for Complex Quantum Systems and Fachbereich Physik, Freie Universität Berlin, 14195 Berlin, Germany}

\author{N.~Paradiso}\email{nicola.paradiso@physik.uni-regensburg.de}
\author{C.~Strunk}
\affiliation{Institut f\"ur Experimentelle und Angewandte Physik, University of Regensburg, 93040 Regensburg, Germany}
%


\begin{abstract}	

Two-dimensional arrays of ballistic Josephson junctions are important as model systems for synthetic quantum materials. Here, we investigate arrays of multiterminal junctions which exhibit a phase difference $\varphi_0$ at zero current. 
When applying an in-plane magnetic field we observe nonreciprocal vortex depinning currents.
We explain this effect in terms of a ratchet-like pinning potential, which is induced by spontaneous supercurrent loops.
Supercurrent loops arise in multiterminal $\varphi_0$-junction arrays as a consequence of next-nearest neighbor Josephson coupling. 
Tuning the density of vortices to commensurate values of the frustration parameter results in an enhancement of the ratchet effect. In addition, we find a surprising sign reversal of the ratchet effect near frustration 1/3. Our work calls for the search for novel magnetic structures in artificial crystals in the absence of time-reversal symmetry.

\end{abstract}

\maketitle

\section{Introduction}

\begin{figure*}[tb]
\centering
\includegraphics[width=\textwidth]{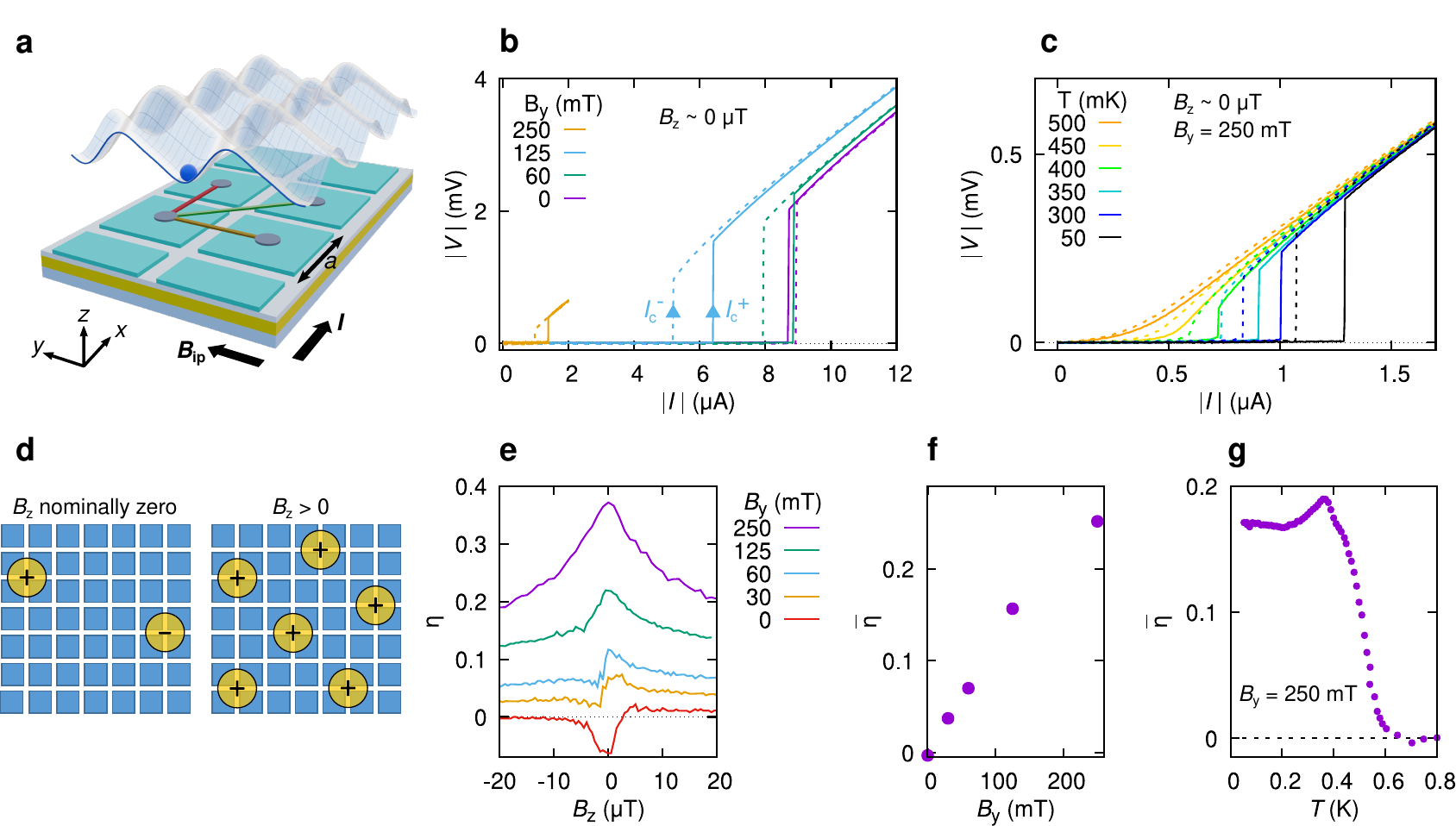}
\caption{\textbf{Nonreciprocal transport in JJAs.}
\textbf{a,} Two-dimensional square Josephson junction array (JJA) with surface plot showing the vortex pinning potential $U(x,y)$, the bias current $I$, and the applied in-plane magnetic field $B_\text{ip}$. The JJA is modeled using both  (red, orange) neighbor and next-nearest (green) neighbor Josephson couplings between superconducting islands. The lattice constant of  the array is $a = 500$~nm.
\textbf{b}, Current-voltage characteristics of the 2D JJA for different magnitude of the in-plane magnetic field. The current is always swept starting at zero absolute value. For the negative branches of the $V(I)$ curves (dotted lines) we plot the absolute values of current and voltage. The out-of-plane magnetic field is close to zero and the temperature is $T\sim 40$~mK.
\textbf{c}, Current-voltage characteristics for different temperatures, measured at an in-plane field $B_y = 250$~mT.
\textbf{d}, Sketch of vortex configuration at nominally zero out-of-plane magnetic field (left) and positive  out-of-plane magnetic field (right).
\textbf{e}, Rectification efficiency $\eta(B_z)$ for $\theta = -90^\circ$ and gate voltage $V_g = 0.5$~V, for different values of $B_\text{ip}$.
\textbf{f}, Average of $\eta$ (labeled as $\bar{\eta}$) in the range $|B_z| < 20$\,$\mu$T  as a function of $B_\text{ip}$, extracted from the data shown in panel.
\textbf{g}, Temperature dependence of $\bar{\eta}$ for $B_\text{ip} = 250$~mT, $\theta = -90^\circ$ and gate voltage $V_g = 0.5$~V. For $T > 400$~mK there is no well-defined jump in the $V(I)$ curves which can be used to define a critical current. Therefore, we use a threshold $V_c = 30$~$\mu$V to define the critical current.
}
\label{fig:Fig1SR}
\end{figure*}

The magnitude of the maximum supercurrent in Josephson junctions is independent of the  polarity of the phase bias when 
time-reversal \textit{or} space-inversion symmetry are present. If both symmetries are lifted, the current-phase-relation (CPR) is no longer odd under inversion of the phase bias. The first experimental evidence for this asymmetry in an individual 
junction was the discovery of the anomalous Josephson effect, i.e., a finite phase shift $\varphi_0$ of the CPR. Such $\varphi_0$-junction behavior has  been demonstrated in systems with large spin-orbit interaction (SOI)~\cite{Szombati2016,Assouline2019,Mayer2020b,Dartiailh2021,Haxell2023}. More recently, similar devices have also featured  nonreciprocal critical currents, referred to as Josephson diode effect~\cite{Ando2020, Baumgartner2022, Pal2022, Jeon2022, Turini2022, Ghosh2024, Costa2023, Lotfizadeh2024, Banerjee2023phase}. As shown in Ref.~\cite{Reinhardt2024}, the anomalous $\varphi_0$-shift and Josephson diode effect can coexist in the same device. 


With the advent of gate-tunable Josephson junction arrays (JJAs) in super-/semiconducting heterostructures~\cite{Boettcher2018}, the natural question arises of how 2D arrays of $\varphi_0$-junctions behave. Already in absence of  $\varphi_0$-shifts, JJAs are established as an important playground in condensed matter physics. They enable the study of fundamental properties of 2D superconductors in a highly controllable fashion~\cite{Boettcher2018} and have been important as model systems for many-body phenomena such as the Berezinski-Kosterlitz-Thouless transition~\cite{Newrock2000,Resnick1981,Abraham1982,vanWees1987,Martinoli2000,Cosmic2020}, quantum phase transitions~\cite{vanderZant1996,Fazio2001,Ikegami2022,Boettcher2018}, phase locking and synchronized emission~\cite{Benz1991, Barbara1999}, and macroscopic quantum effects~\cite{vanderZant1991, Elion1993, Delsing1994}. In perpendicular magnetic fields, the resistive state of the arrays is controlled by vortex dynamics. The vortex depinning current displays striking commensurability effects  at fractional values of the frustration $f=\Phi/\Phi_0$~\cite{Poccia2015, Lankhorst2018,Penner2023, Boettcher2023}, where $\Phi$ is the magnetic flux threading a plaquette and $\Phi_0$ is the superconducting flux quantum. 

Introducing and tuning the $\varphi_0$-shift by an in-plane magnetic field provides a novel knob for controlling JJAs. For the simplest case of square arrays with nearest neighbor Josephson coupling only, a uniform $\varphi_0$-shift in all junctions can be gauged away and has no  experimentally observable consequences. It is an open question, under which conditions the $\varphi_0$-shift has an observable impact on the transport characteristics of JJAs.

Here, we report on two-dimensional square arrays made of $\varphi_0$-junctions. Applying an in-plane magnetic field with a component perpendicular to the current, we observe nonreciprocal vortex depinning currents. We make use of this new type of supercurrent diode effect to probe the impact of $\varphi_0$-shifts on 2D JJAs.  At frustrations $f\ll1$, the nonreciprocity persists up to fairly large in-plane fields and temperatures
. This effect is explained in terms of a field-tunable, ratchet-like shape of the vortex pinning potential, which we deduce from a minimal model of the JJA with both nearest and next-nearest neighbor Josephson couplings with anomalous $\varphi_0$-shifts.
Vortex nonreciprocity also occurs at fractional $f$, with inverted sign for $f\simeq1/3$. 

In our devices, the 2DEG is located in a shallow InGaAs/InAs/InGaAs quantum well, whose characteristics are described in detail in the Supplementary Information. Superconductivity in the 2DEG is introduced by proximity to an epitaxially-grown Al film~\cite{Shabani2016}. Using electron beam lithography followed by selective wet-etching of the Al film, we define a square array of 200$\times$200 aluminum islands separated by 100~nm wide gaps where the aluminum is removed. The islands are 400$\times$400~nm$^2$
in size. A global top-gate allows us to control the electron density in the weak links.

For $\Phi\ll\Phi_0$, the square JJAs (lattice constant $a=500$~nm) feature an intrinsic vortex pinning potential --also called egg-crate potential~\cite{Newrock2000}-- $U(x,y)\approx E_B[ \cos (2\pi x/a) + \cos (2\pi y/a)]$ 
with minima of the potential located near the corners of the superconducting islands, as illustrated in Fig.~\ref{fig:Fig1SR}\textbf{a}. When applying a transport current with density $\vec{j}$, vortices experience a Lorentz force $\vec{F}_L = \Phi_0\vec{j} \times \vec{z}$  perpendicular to the applied current. Depinning of vortices occurs when the Lorentz force exceeds the maximal gradient of the pinning potential. For a single vortex in a square array the depinning current density is $j_c = 2\pi E_B / a\Phi_0$, which is approximately 10\% of the critical current density of the individual junctions 
\cite{Rzchowski1990,Newrock2000,Fazio2001}. Thus, in 2D JJAs, as soon as vortices are present, the measured critical current is determined by their depinning threshold and not by the individual junction critical current.

\begin{figure*}[htb]
\centering
\includegraphics[width=\textwidth]{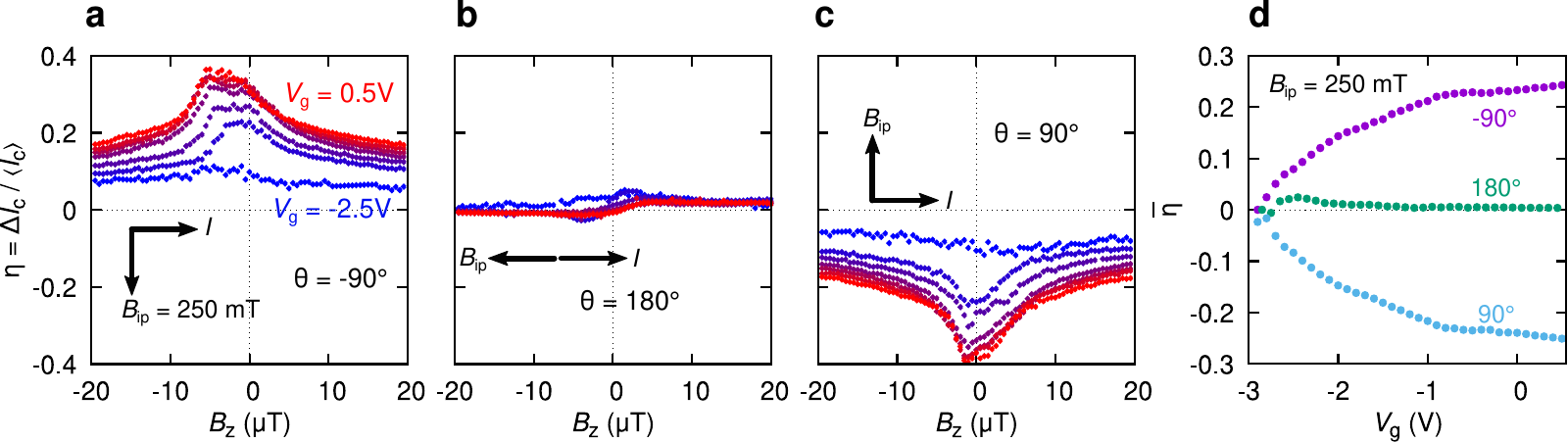}
\caption{\textbf{Tunability of rectification.}
\textbf{a-c}, Rectification efficiency $\eta = \Delta I_\text{c} / I_\text{c,mean}$ for different orientations (black arrows) of the in-plane field, measured at $B_\text{ip} = 250$~mT and temperature $T \sim 40$~mK kept constant during the entire measurement. The color corresponds to different values of the gate voltage $V_g$, varied in steps of 0.5~V from -2.5~V to 0.5~V.  
\textbf{d}, Average of $\eta$ (labeled as $\bar{\eta}$) in the range $|B_z| < 20$~$\mu$T at $B_{\text{ip}} = 250$~mT as a function of gate voltage for different orientations of the in-plane magnetic field.
}  
\label{fig:Fig2SR}
\end{figure*}

\section{Results}
Figures~\ref{fig:Fig1SR}\textbf{b-g} show dc transport data in the regime of very dilute vortices (frustration $f\ll 1$, where $f = B_z/B_0$ with $B_0 = \Phi_0/a^2 = 8.2$~mT). Panel \textbf{b} shows $V(I)$-characteristics for different magnitudes of the in-plane magnetic field which is applied perpendicular to the direction of the transport current. A large difference between positive and negative critical currents is observed for non-zero in-plane fields. While the critical current is strongly reduced at the largest in-plane field of $B_y = 250$~mT, a sizable asymmetry is still visible. The temperature dependence of $V(I)$-characteristics at $B_y=250$~mT is shown in panel \textbf{c}. The nonreciprocity of the $V(I)$ curves persists with increasing temperature until a well-defined critical current is no longer observable at $T\sim 450$~mK. 
To quantify the degree of nonreciprocity, panel \textbf{e} shows the rectification efficiency $\eta \equiv (I_c^+-|I_c^-|)/\langle I_c\rangle$ (with $\langle I_c\rangle\equiv (I_c^++|I_c^-|)/2$) as a function of out-of-plane field $B_z$ for different magnitudes of the in-plane field $B_y$. 
The central observation is that both the critical current and the rectification efficiency $\eta$ are even functions of $B_z$ and display a sharp --roughly 10 $\mu$T-wide-- peak near $B_z$. Even at nominally $B_z=0$ the critical current density is about ten times less than the Josephson junction critical current density. Thus we conclude that the observed switching to the normal state is signaling the depinning of Josephson vortices. The fact that Josephson vortices are present at nominally $B_z=0$ is not too surprising. 
A single vortex in our array with size $100 \times 100$\,$\mu$m$^2$ corresponds to a field of $200$\,nT.
The out-of-plane field that we use to compensate the misalignment of the in-plane field has always a finite inhomogeneity in $(x,y)$.  Even a tiny inhomogeneity (sub-$\mu$T over hundreds of micrometers) is sufficient to obtain
vortices and antivortices somewhere in the array, even though $B_z$ averaged over the array can be accurately set to zero, see sketch in Fig.~\ref{fig:Fig1SR}\textbf{d}, left. When discussing our theory model, we shall demonstrate that the spin-orbit induced  rectification efficiency has the same sign for vortices and antivortices (implying that $\eta(B_z)$ is even). This is a peculiar feature of spin-orbit induced ratchets which contrasts with the behavior of ratchets demonstrated so far, which are obtained by real-space asymmetry of the potential.


The rectification efficiency $\eta$ as a function of $B_z$ is shown in Fig.~\ref{fig:Fig1SR}\textbf{e} for different values of the in-plane field $B_y$.
Averaging $\eta$ over the field range $|B_z| < 20$\,$\mu$T, ($f < | 0.0024|$) results in the plot shown in panel \textbf{f} which shows that $\bar{\eta}(B_y)$ 
increases linearly up to $125$\,mT and continues to increase with lower slope up to $250$\,mT.
As a function of temperature $\bar{\eta}$ shows no decrease until the temperature is increased above $T\sim 400$~mK, as shown in panel \textbf{g}. Both observations confirm that the observed rectification efficiency is not related to the single Josephson junction diode effect~\cite{Baumgartner2022}: in fact, the Josephson diode effect in similar Rashba systems always shows a maximum at roughly 100~mT, followed by a rapid decay~\cite{Baumgartner2022,Costa2023,Reinhardt2024,schiela2025}. Also, the suppression of higher harmonics (and thus of the diode effect) in such systems is already substantial above 100~mK. On the other hand, the $\varphi_0$-shift is expected to grow monotonically and to be nearly constant in temperature~\cite{Reinhardt2024}. As we shall discuss below, it is precisely the $\varphi_0$-shift which induces the nonreciprocal vortex depinning.


 

If the anomalous Josephson effect is the key for the rectification, we expect a marked dependence of $\eta$ both on the angle between in-plane field and current, and on the gate voltage controlling the Rashba coefficient and Fermi level.
Figures~\ref{fig:Fig2SR}\textbf{a-c} show $\eta$ as a function of out-of-plane field $B_z$ and gate voltage for different orientations of $\vec{B}_{\text{ip}}$. The rectification is suppressed when the in-plane field is aligned in the direction of the bias current. Moreover, changing the sign of $B_y$ changes the sign of $\eta$. As a function of gate voltage, the averaged rectification efficiency $\bar{\eta}$ monotonically increases with gate voltage (panel \textbf{d}), showing that the nonreciprocity is enhanced by a larger electron density in the weak-link. Also the gate dependence of $\eta$ has a similar behavior as that of $\varphi_0$ reported in similar systems~\cite{Reinhardt2024, Lotfizadeh2024}.
The magnitude and sign of the nonreciprocity are reproduced on the nominally identical device B. The measurements on device B are presented in the Supplementary Figures~S7\textbf{d,e}. We have also performed measurements on a JJA (device C) with larger lattice constant $a=1.1$~$\mu$m. In addition, the InAs quantum well of device C has a mean-free path of $l_\text{max}<270$~nm, which is much lower compared to the mean-free path $l_\text{max} \sim 700$~nm of the quantum well used for devices A and B. The measurements of device C are presented in Figure S10 of the Supplemental Information. Our main observation is that the nonreciprocity is much weaker in device C, owing to the larger lattice constant compared to that in devices A and B.

Let us summarize our observations so far:
The switch to the normal state is observed at current densities which are one order of magnitude below the expected Josephson critical current density. E.g., at low in-plane fields $B_\text{ip} \lesssim 100$~mT the critical current density is $j_c \approx  90$~nA/$\mu$m, while in the case of a ballistic junction, the expected Josephson critical current would be $j_{c,\text{JJ}} = \pi \Delta^* / (e R_N a) \sim 1.9$~$\mu$A/$\mu$m. Here, we use the measured value $R_N \sim 500$~$\Omega$ 
and the estimated induced gap  $\Delta^* = 130$~$\mu$eV~\cite{baumgartner2020}.
Thus, the observed critical currents agree reasonably with the expected vortex depinning currents.
The vortex depinning is nonreciprocal if an in-plane field is applied perpendicular to the current. This diode effect shows a gate, temperature, and in-plane field dependence similar to that reported for the $\varphi_0$-shift. Finally, $\eta(B_z)$ is even in the out-of-plane field $B_z$.


\begin{figure*}[tb]
\centering
\includegraphics[width=\textwidth]{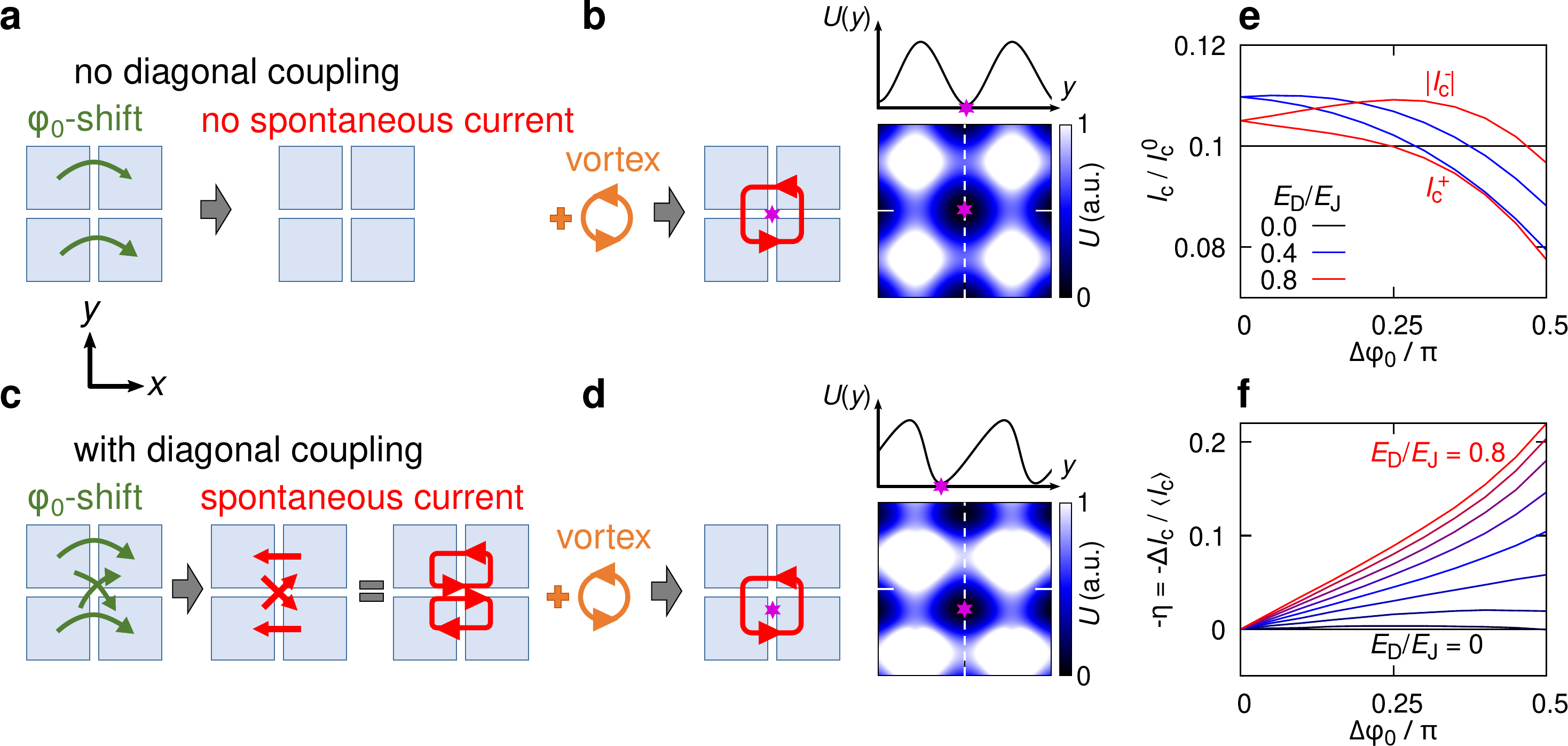}
\caption{\textbf{Model of the ratchet effect.}
\textbf{a,} JJA with nearest neighbor coupling only, with no spontaneous supercurrent in the ground state. 
\textbf{b,} Sketch of the Josephson currents for a vortex in absence of diagonal coupling. The Josephson current distribution remains four-fold symmetric. A color plot of the vortex pinning potential is shown on the right.
Vortices driven in the $y$-direction experience a sinusoidal pinning potential $U(y)$, as sketched above the color plot of $U(x,y)$.
\textbf{c,} Anomalous phase shifts $\varphi_0^x > \varphi_0^{\text{diag}}$ and persistent currents for the case with diagonal coupling.
\textbf{d,} Sketch of the Josephson currents for a vortex in the case with diagonal coupling. The corresponding vortex pinning potential $U(x,y)$ is skewed. Vortices driven in the $y$-direction experience a ratchet-like potential $U(y)$, as sketched above the color plot of $U(x,y)$ [$U(y)$ shown with exaggerated skewness for better visibility].
\textbf{e,} Numerical simulation of vortex depinning currents. The phase shift parameter $\Delta \varphi_0$ is the difference between horizontal and diagonal phase shifts: $\Delta \varphi_0 = \varphi_0^x - \varphi_0^{\text{diag}}$. $E_D/E_J$ is the ratio between diagonal and non-diagonal Josephson couplings.
\textbf{f,} Resulting rectification efficiency $\eta = \Delta I_c / \langle I_{c} \rangle$
 extracted from the numerical simulation of positive and negative depinning currents. The diagonal coupling is changed in steps of $0.1$ from $0$ to $0.8$. }

\label{fig:Fig3SR}
\end{figure*}

\begin{figure*}[htb]
\centering
\includegraphics[width=\textwidth]{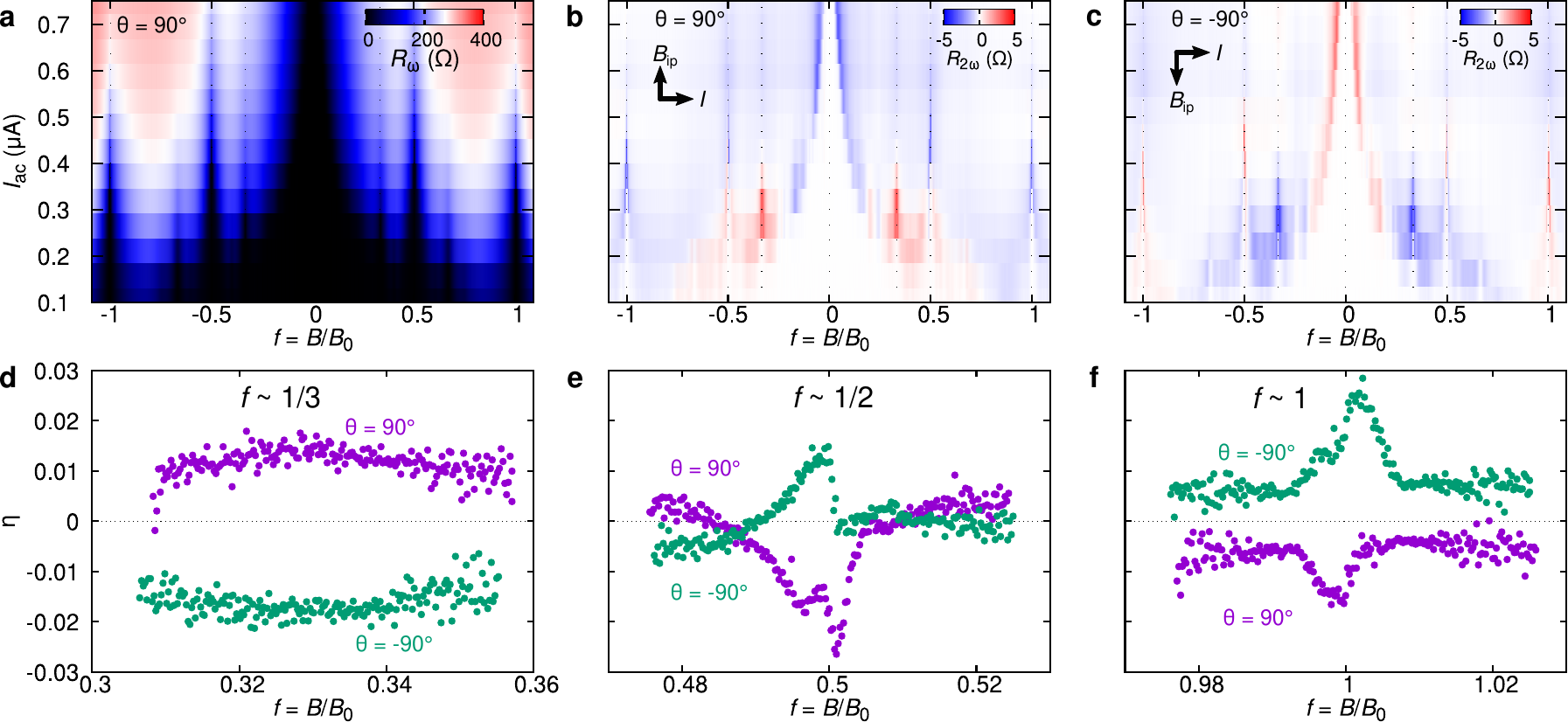}
\caption{\textbf{Ratchet effect at larger frustration}
\textbf{a-c}, First harmonic $R_\omega = V_\omega / I_{ac}$ and second harmonic $R_{2\omega} = V_{2\omega} / I_{ac}$ of the resistance measured as a function of frustration and ac bias current. The in-plane field $B_\text{ip} = 125$~mT is applied perpendicular to the direction of current.
\textbf{d-f}, Rectification efficiency $\eta$ around commensurate fields $f = 1/3$, $f = 1/2$, 
and $f=1$ obtained from standard $V(I)$ transport measurements.
In all plots $T \sim 40$~mK.
}

\label{fig:Fig4SR}
\end{figure*}

As mentioned above, a nonreciprocal effect is unexpected for a 2D array of $\varphi_0$-junctions when modeling the JJA with a 2D XY model with nearest-neighbor coupling, where islands are linked by junctions in horizontal and vertical directions. However, this model neglects the strong multi-terminal character of the junctions in our devices A and B. The magnitude and range of the coupling between islands is ultimately determined by the spatial extent of the Andreev bound states in the proximitized electron gas. The important length scales are the coherence length $\xi_0 = \frac{\hbar v_F}{\pi \Delta^*}$ and the mean free path $l_e$. From the characterization data (which can be found in the Supplement of \cite{Reinhardt2024}) we deduce a lower bound of 800~nm for $\xi_0$ (assuming $\Delta^{\ast}=130$~$\mu$eV and $v_F=5\cdot 10^5$~m/s) and 250~nm for $\ell$. Therefore, we can estimate the spatial extent of the Andreev bound states under the Al islands $\xi=\sqrt{\hbar D/\Delta^{\ast}}=\sqrt{\xi_0 \ell/2}$ to be at least 300~nm, which is comparable with the island size. 
To link the observed nonreciprocal vortex depinning current to the anomalous phase shifts, we transcend the standard XY model with only nearest neighbor couplings and take the multi-terminal character~\cite{Gupta2023,VirtanenPRL2024} of the junctions into account.  Microscopically, 
Andreev bound states (ABS) in the weak links connect not only nearest neighbor Al islands, but
to a lesser extent also next-nearest neighbors as indicated in Fig.~\ref{fig:Fig1SR}\textbf{a} (green line).
With in-plane magnetic fields along the $\hat{y}$ direction, the diagonal junctions also exhibit a non-zero phase shift $\varphi_0^{\text{diag}}$, which in general differs from the phase shift $\varphi_0^x$ of the junctions in the $x$-direction. The additional diagonal couplings render the ground state frustrated even without perpendicular magnetic field: The sums of $\varphi_0$-shifts around closed loops no longer cancel. To maintain fluxoid quantization within the plaquettes, spontaneous supercurrents emerge in the ground state, see Fig.~\ref{fig:Fig3SR}\textbf{c}. The current configuration in the ground state has an (up-down) reflection symmetry about the current axis. This symmetry is broken when a vortex is added to the array, as shown in Fig.~\ref{fig:Fig3SR}\textbf{d}. The vortex pinning potential $U(x,y)$ shown on the right side of Fig.~\ref{fig:Fig3SR}\textbf{d} no longer exhibits fourfold rotational symmetry and $U(x=0,y)$ exhibits a ratchet-like dependence (see the Supplementary Information for calculation details).

The importance of the diagonal couplings is further emphasized by computing the vortex depinning currents. We minimize the free energy of the JJA with a single vortex located within the array. The depinning current is obtained as the largest bias current, for which the vortex remains localized (for details, see the Supplemental Information). The results are plotted in Fig.~\ref{fig:Fig3SR}\textbf{e} as a function of the difference $\Delta \varphi_0 = \varphi_0^x - \varphi_0^{\text{diag}}$ between horizontal and diagonal phase shifts. The phase shift affects the depinning currents only when the coupling $E_D$ between next-nearest neighbors is turned on and results in a nonreci\-procal depinning current. The corresponding diode efficiencies $\eta \equiv (I_c^+-|I_c^-|)/\langle I_c\rangle$ (with $\langle I_c\rangle\equiv (I_c^+ + |I_c^-|)/2$) are shown in Fig.~\ref{fig:Fig3SR}\textbf{f} and reach up to $20$~\% for $\Delta\varphi_0 = \pi/2$ and $E_D = 0.8$, comparable to experimental values.
Interestingly, in the limit of small frustrations, the nonreciprocal vortex dynamics can be mapped theoretically to the phase dynamics of a single Josephson diode \cite{Steiner2023}. The motion of an individual vortex along the $\hat y$ direction in the pinning potential is governed by the Langevin equation 
\begin{equation}
    -\partial_y U - \alpha \dot y - \gamma \Phi_0 I  = f_y
    \label{eq:langevin}
\end{equation}
(see Supplemental Information for details, including the geometrical factor $\gamma$). The first term describes the pinning force, the second the friction, while the third term corresponds to the Lorentz force exerted by the bias current $I$ in the $\hat x$ direction. The Langevin force $f_y$ has zero average and correlator $\langle f_y(t)f_y(t')\rangle = 2\alpha k_BT \delta(t-t')$. Equation (\ref{eq:langevin}) maps onto the Langevin equation for the phase difference of a single resistively shunted Josephson junction, with the vortex position becoming the phase difference and the ratchet-like vortex pinning potential turning into the asymmetric current-phase relation.

At this point, we are also able to explain the even dependence of $\eta$ on $B_z$. The key idea is that for an antivortex the potential induced by the spontaneous supercurrents is mirrored in $y$ compared to that for a vortex. This is illustrated in Fig. S4 of the Supplementary Information. Since also the direction of the Lorentz force is opposite, the sign of $\eta$ remains the same. This behavior is markedly different from the engineered ratchet effect obtained with a real-space asymmetric pinning potential. In that case the potential is the same for vortices and antivortices, so that the resulting rectification efficiency is opposite.
A more detailed discussion of this important point can be found in the Supplement.

Beyond the limit of dilute vortices, the rectification coefficient also exhibits interesting behavior at larger frustrations.
We probe the vortex dynamics by applying a low frequency ac current bias with amplitude $I_{ac}$ while measuring the first and second harmonic of resistance using digital lock-in amplifiers. This provides a convenient and fast method to measure nonreciprocal response~\cite{Wakatsuki2017,Ando2020}.  The second harmonic of the ac response $R_{2\omega} \equiv V_{2\omega}/I_{ac}$ becomes non-zero when the amplitude of the ac current is in the rectification window ($|I_c^-| < I_{ac} < I_c^+$) where $V(I) \ne -V(-I)$. 
A simulation of $R_\omega(I_{ac})$ and $R_{2\omega}(I_{ac})$ for a current-voltage characteristic with nonreciprocal critical currents is provided in the Supplementary Information.
The $R_{2\omega}$ measurement provides better signal-to-noise ratios compared to dc measurements of IV-characteristics. 
Fig.~\ref{fig:Fig4SR}\textbf{a} shows the first harmonic $R_\omega \equiv V_\omega / I_{ac}$ as a function of $f$ and $I_{ac}$, measured for an in-plane field $B_\text{ip} = 125$~mT at an angle of $\theta=90^\circ$ with the applied current $\vec{I}$. A nonzero resistance is observed when $I_{ac}$ exceeds the vortex depinning current, which strongly depends on the applied out-of-plane field. We observe pronounced maxima of the depinning current for the commensurate values of frustration $f = \pm 1/3, \pm 1/2, \pm 2/3, \pm 1$, where vortices form ordered patterns and pinning is strongly increased \cite{Tinkham1983, Newrock2000}.
$R_{2\omega}$ is shown in Fig.~\ref{fig:Fig4SR}\textbf{b,c} for 
$\theta = \pm 90^\circ$. Peaks of $R_{2\omega}$ with the same sign (blue color in Fig.~\ref{fig:Fig4SR}\textbf{b}, red in Fig.~\ref{fig:Fig4SR}\textbf{c})  are observed at $f=0$, $f=1/2$, and $f=1$.
A pronounced peak of $R_{2\omega}$ with reversed sign (red color in Fig.~\ref{fig:Fig4SR}\textbf{b}, blue in Fig.~\ref{fig:Fig4SR}\textbf{c}) is found at $f=1/3$. The reversed sign can be found in a wider region of frustrations around $f=1/3$, approximately given by $0.2 \lesssim f \lesssim 0.5$. The sign reversal can be reproduced at a lower in-plane field of $60$~mT (Fig.~S5 of Supplementary Information). These findings are substantiated when extracting $\eta$ from IV-measurements for frustrations close to the commensurate values $f=1/3, 1/2$ and $1$, see Fig.~\ref{fig:Fig4SR} \textbf{d-f}. 
\section{Discussion}
We would like to emphasize a key difference between our \textit{magnetochiral ratchet} effect and the vortex ratchet effect reported in previous experiments. So far, a ratchet-like pinning potential was obtained by breaking the real-space symmetry of the system, for example by asymmetrically fabricated pinning sites~\cite{Gillijns2007, Shalom2005}. Here, the array remains four-fold symmetric ($D_4$ symmetry) and the symmetry of the pinning potential is reduced to that of a ratchet ($D_1$) by the combination of SOI, Zeeman field, and diagonal couplings. In our magnetochiral ratchets, the rectification efficiency is the same for vortices and antivortices, while it is opposite for asymmetrically fabricated pinning sites \cite{Gillijns2007, Shalom2005}.

A change of sign of the vortex ratchet effect has been previously reported in arrays with asymmetric potential modulation~\cite{Shalom2005, Marconi2007, Villegas2003, Silva2006, Gillijns2007, Lu2007}.
In our case, however, the physics is different: the change of sign we observe is an emergent property of symmetric and periodic arrays, which is evidently related to the particular vortex patterns at commensurate frustration values.

In conclusion, we observe a magnetochiral vortex ratchet effect in 2D arrays of $\varphi_0$-junctions. The multiterminal character of our junctions introduces competing $\varphi_0$-shifts, leading to frustration and the emergence of spontaneous supercurrent loops in the ground state. 
These spontaneous currents are ultimately responsible for the nonreciprocity in the depinning current. Our experimental results for $f\ll 1$ are in nice agreement with a minimal model for a 2D square array with nearest and next-nearest-neighbor Josephson couplings, while the sign change at $f=1/3$ remains an interesting open question.
\section{Materials and Methods}
Samples are fabricated starting with a heterostructure which is grown by molecular beam epitaxy.
The full layer sequence and basic characterization of this heterostructure are presented in the Supplemental Information of our previous work \cite{Reinhardt2024}. The top-most layers are a $5$~nm film of Al, a $10$~nm In$_{0.75}$Ga$_{0.25}$As barrier, the $7$~nm thick InAs layer, which hosts the 2DEG, and a $4$~nm bottom barrier of In$_{0.75}$Ga$_{0.25}$As. The devices are structured using standard nanofabrication techniques. Deep etching of the heterostructure is performed using a phosphoric acid solution. The arrays are definded using electron beam lithography and selective etching of the Al film with Transene D.

Transport measurements are performed in a dilution refrigerator with a base temperature of $40$~mK. All measurement lines are filtered at room temperature using Pi-type LC filteres with a cutoff frequency of $10$~MHz. Below the mixing chamber we use resistive coaxial wires which have a large attenuation at frequencies above $100$~MHz. More details of the setup can be found in the Supplemental Information of our previous work \cite{Reinhardt2024}.

The configuration of vortices in the array and the superconducting contacts will in general depend on the history of the out-of-plane magnetic field. Field-cooling is required to obtain an equilibrium configuration of vortices.
The data presented in the main text has been obtained with the following field-cooling procedures:
\begin{itemize}
    \item Fig.~1 \textbf{e}: field-cooling at every value of $B_z$
    \item all other panels of Fig.~1 and Fig.~2: field-cooling in nominally zero out-of-plane field before the measurement
    \item Fig.~4 \textbf{a-c}: no field cooling
    \item Fig.~4 \textbf{d-f}: field-cooling at the commensurate fields ($f=1/3$, $1/2$, and $1$) before the measurement
\end{itemize}

The full description of theoretical methods is provided in the Supplemental Information.

\begin{acknowledgments}
We thank S.~Vaitiekenas for fruitful discussions.
Work at Universit\"at Regensburg was funded by the EU’s HORIZON-RIA Programme under Grant No.\ 101135240 (JOGATE), and by Deutsche Forschungsgemeinschaft (DFG, German Research Foundation) through Project-ID 314695032—SFB 1277 (Subproject  B08). Research at Freie Universit\"at Berlin was supported through Collaborative Research Center (CRC) 183 (project C03) of the Deutsche Forschungsgemeinschaft and the Einstein Research Unit on Quantum Devices. Research at Yale University was supported by the Office of Naval Research (ONR) under award number N00014-22-1-2764 and by the NSF Grant No.\ DMR-2410182. L.I.G.\ thanks Freie Universit\"at Berlin for hosting him as a Mercator fellow within CRC 183.

\end{acknowledgments}
\vspace{2mm}





\bibliography{biblio}

\clearpage
\newpage
\beginsupplement

\onecolumngrid
\begin{center}
\textbf{\Large Supplementary Information} 

\vspace{0.5cm}

\noindent \textbf{\large Spontaneous supercurrents and vortex depinning in two-dimensional arrays of $\varphi_0$-junctions}
\end{center}

\section{Theoretical methods} 

\subsection{Model} 

We model the Josephson junction array as a $N \times N$ square lattice of  superconducting islands coupled via Josephson junctions.
Each island (labeled by site indices $i,j$ in the $x$ and $y$ directions) 
is described by the phase $\varphi_{i,j}$ of the superconducting order parameter. 
To describe the experiment, we transcend standard treatments in two ways:
\begin{itemize}
    \item In addition to Josephson couplings between nearest neighbors along the $x$- and $y$-directions (strength $E_J$), we also include next-nearest neighbor  couplings along the diagonals of the square lattice (Josephson energy $E_D$). This is motivated by the experimental geometry of square superconducting islands separated by narrow strips of 2DEG, which facilitates Josephson coupling along the diagonal direction.  

    \item We account for the $\varphi_0$-junction behaviour of the horizontal and diagonal junctions, which arises from the interplay of spin-orbit coupling and in-plane magnetic field. Taking the magnetic field along the $y$-direction, the current-phase relations of both, the horizontal and diagonal junctions are characterized by a phase offset. Since horizontal and diagonal junctions enclose different non-zero angles with the in-plane field, their phase offsets are different and denoted by $\varphi^x_0$ and $\varphi^\mathrm{diag}_0$, respectively.
\end{itemize}

Our model Hamiltonian takes the form
    \begin{align}
        H &= H_0 + H_D, \label{eq: Hamiltonian} \\
        H_0 &= -E_J\sum_{i=1}^{N-1}\sum_{j=1}^{N} \cos(\varphi_{i+1,j} - \varphi_{i,j} -\varphi^x_0)
    -E_J\sum_{i=1}^{N}\sum_{j=1}^{N-1} \cos(\varphi_{i,j+1} - \varphi_{i,j} ), \label{eq: Hamiltonian0} \\
    H_D &= -E_D\sum_{i,j=1}^{N-1} \left[\cos(\varphi_{i+1,j+1}-\varphi_{i,j} - \varphi^\mathrm{diag}_0) +\cos(\varphi_{i+1,j} - \varphi_{i,j+1} - \varphi^\mathrm{diag}_0)\right],
    \label{eq: Hamiltoniand}
    \end{align}
where $H_0$ accounts for the horizontal and vertical junctions and $H_D$ for the diagonal couplings.

In the absence of the diagonal couplings, the phase offsets $\varphi^x_0$ are inconsequential, as they do not change the sum of gauge-invariant phase differences around any of the plaquettes. 
When including the diagonal junctions, there are additional triangular plaquettes. The phase offsets modify the sum of gauge-invariant phase differences around these plaquette by $\pm (\varphi^x_0 - \varphi^\mathrm{diag}_0)$ as illustrated in Fig.\ \ref{fig: flux,groundstate}(a). 
We can simplify the model and reduce the number of independent parameters by noting that a configuration with zero phase offsets for the diagonal junctions and offsets $\Delta\varphi_0 = \varphi^x_0 - \varphi^\mathrm{diag}_0$ for the horizontal junctions is gauge equivalent. 

\begin{figure*}[t!]
    \includegraphics[scale = 0.55]
    {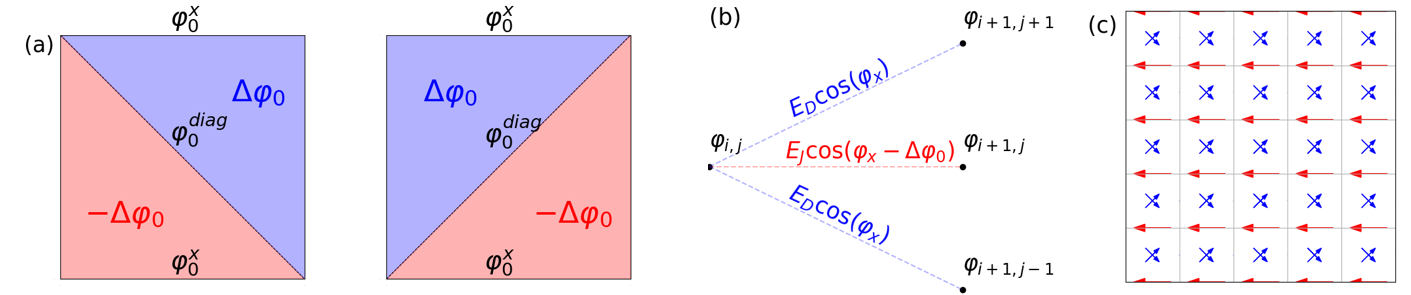}
    \caption{Model Hamiltonian with $\varphi_0$-junctions and diagonal couplings. (a) Phase configuration inside of a plaquette. Due to the diagonal couplings and the phase offsets of the junctions, the system is effectively subject to a transverse magnetic flux $ \Delta\varphi_0 = \varphi^x_0-\varphi^\mathrm{diag}_0$ in the blue region and $-\Delta\varphi_0$ in the red region. (b) Illustration of the ground-state calculation. Only an elementary cell involving four superconducting islands need to be considered due to translational symmetry in the $x$- and $y$-directions. (c) Current configuration of the ground state. The arrows indicate the currents between the respective nodes. The length of the arrows indicates the magnitude of the current, with the critical current corresponding to an arrow connecting the nodes. Parameters: $E_D=0.5E_J$, $\Delta\varphi_0 = \pi/2$.}
    \label{fig: flux,groundstate}
\end{figure*}

The phase offsets modify the ground-state phase configuration $\varphi^{\text{GS}}_{i,j}$. In the limit of large $N$, we can assume that all phases along a column are aligned as there is no phase bias for the vertical junctions. (Strictly speaking, this assumes that $E_D<E_J$, which is clearly satisfied in experiment.) Denoting the phase difference between neighboring columns by 
$\varphi_x$, we then have
$\varphi^{\text{GS}}_{i,j} = i \varphi_x$. The coupling energies of an island with its three (nearest and next-nearest) neighbors to the right [see Fig.\ \ref{fig: flux,groundstate}(b) for an illustration] is uniform across the entire array, so that the ground-state phase configurations  minimizes 
\begin{equation}
    h(\varphi_x) = -E_J \cos(\varphi_x - \Delta\varphi_0) -2E_D\cos \varphi_x.
\end{equation}
This yields
\begin{equation}    \varphi_x = \arctan(\frac{E_J\sin\Delta\varphi_0}{E_J\cos \Delta\varphi_0 + 2E_D}),
\end{equation}
so that the ground-state phase configuration can be written explicitly as
\begin{equation}
    \varphi^{\text{GS}}_{i,j} = i\arctan(\frac{E_J\sin\Delta\varphi_0}{E_J\cos\Delta\varphi_0 + 2E_D}).
    \label{eq: groundstate}
\end{equation} 
We have confirmed for various parameters $E_D$ and $\Delta\varphi_0$ that this is consistent with numerical results obtained by minimizing the energy of the entire array. 

The phase configuration implies that the diagonal couplings along with the phase offsets lead to currents flowing in the ground state, namely  
$I_h = (2eE_J/\hbar)  \sin(\varphi_x-\Delta\varphi_0)$ along the horizontal bonds and $I_d = (2e E_D/\hbar) \sin \varphi_x $ along the diagonal bonds. This is illustrated in Fig.\ \ref{fig: flux,groundstate}(c). We note that the currents break the mirror symmetry with respect to the vertical axis as well as time-reversal symmetry, which is conducive for a diode effect. 

\subsection{Vortex potential}

\begin{figure*}
\hspace*{-0.2cm}
    \includegraphics[scale=0.63]{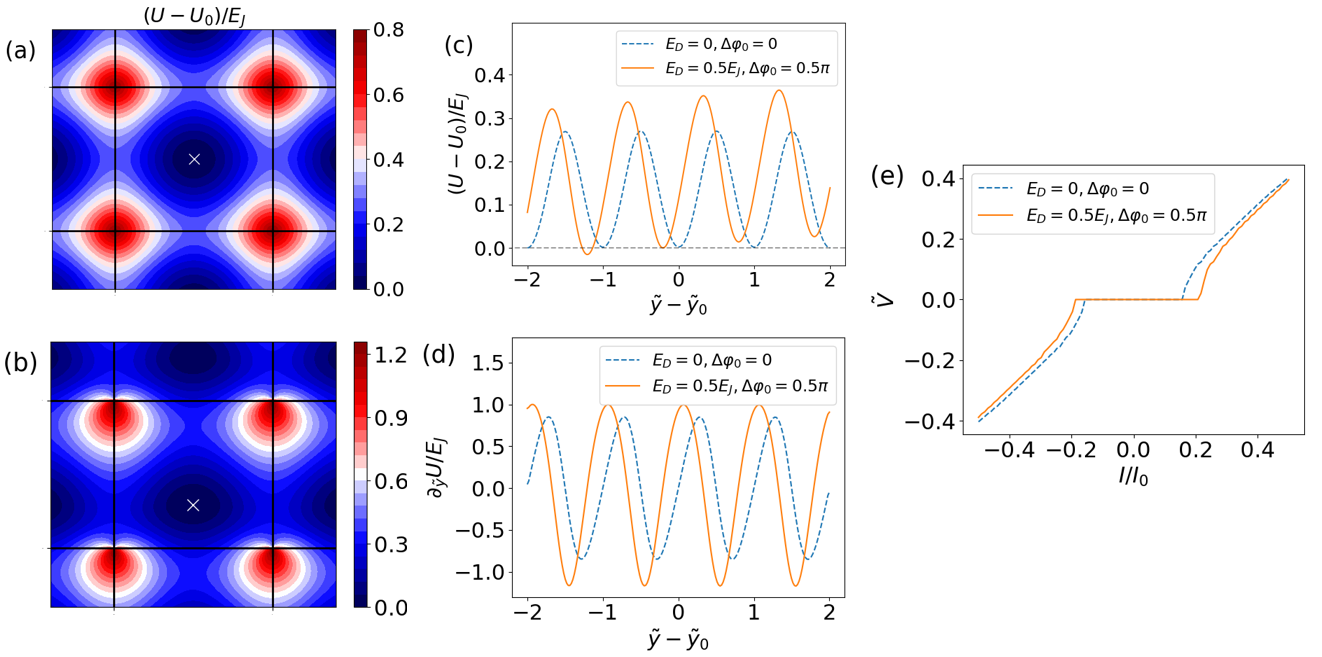}
    \caption{Vortex in a Josephson junction array with $\varphi_0$-junctions and diagonal couplings. (a) Contour plot of the vortex potential $U(x,y)$ with $E_D=0$ and $ \Delta\varphi_0=0$. The black lines indicate the Josephson junction array, the white cross shows the minimum of the potential. (b) Same plot as (a) with $E_D = E_J/2$ and $\Delta\varphi_0 = \pi/2$. (c) Plot of the vortex potential  $U(x_0,y-y_0)$ as a function of $y$, with $(x_0,y_0)$ denoting coordinates of the array center. Parameters corresponding to both (a) and (b). For easier comparison, we subtract a constant $U_0$ corresponding to the potential at the minimum closest to the array center. (d) $y$-derivatives of the potential as a function of $y$ corresponding to (c). (e) $I-V$ curves for both parameter sets calculated from the Langevin equation in Eq.\ \ref{eq:langevin}. The depinning currents correspond to the extreme values in (d) weighted by the reduction parameter $\gamma$. We calculated the voltage for 200 current values between $-I_0/2$ and $I_0/2$. Further parameters: $\tau = 0.05$, $n_{\text{max}} = 2000$.}
    \label{fig: potential,vortex}
\end{figure*}

The experiment suggests that a low density of vortices is present even at nomininally vanishing perpendicular magnetic field. We consider the dynamics of a single vortex with given circulation. 

We first approximate the phase configuration in the presence of a vortex by \cite{Rzchowski1990}
\begin{equation}
    \varphi_{i,j}(x,y) = \varphi^{\text{GS}}_{i,j} + \varphi^v_{i,j}(x,y), 
    \label{eq: vortex phase configuration}
\end{equation}
where
\begin{equation}
    \varphi^v_{i,j}(x,y) = \arctan(\frac{j-y}{i-x}).
\end{equation}
This neglects relaxation effects due to the interplay of the ground-state currents with the vortex configuration. (In the next section, we will go beyond this approximation.)

We can use this ansatz to calculate the potential energy $U$ of the vortex as a function of its position $(x,y)$ (taken to be continuous) through
\begin{equation}
    U(x,y) = H(\{\varphi_{i,j}(x,y)\}).
\end{equation}
Here, $H$ denotes the  Hamiltonian in Eq.\ (\ref{eq: Hamiltonian}).
For $E_D=0$ and $\Delta\varphi_0=0$, we reproduce the results of Ref.\ \cite{Rzchowski1990} as shown in Fig.\ \ref{fig: potential,vortex}(a). The potential minima are located in the center of the plaquettes, while the maxima sit at the superconducting sites. 

Turning on the diagonal couplings as well as the phase offsets leads to significant modifications of the vortex potential, as shown in Fig.\ \ref{fig: potential,vortex}(b) for $E_D=0.5 E_J$ and $\Delta\varphi_0=\pi/2$. The minima are shifted away from the plaquette centers along the $y$-direction. The associated deformation of the potential breaks the mirror symmetry about the $x$-axis. These differences are further illustrated by line cuts of the potential and its derivative as shown in Fig.\ \ref{fig: potential,vortex}(c,d). These cuts are along lines of fixed $x$, with $x$ taken at the center of the plaquette. The diagonal couplings clearly shift the minima of the potential. Moreover, the maximal positive derivative of $U$ along the $y$-direction becomes smaller than the maximal negative derivative, reflecting a tendency of the potential to form a ratchet 
[Fig.\ \ref{fig: potential,vortex}(d)]. In addition, finite-size effects lead to a monotonously changing shift of the potential across the sample, which implies asymmetric upward and downward barriers. 

The ratchet-like vortex potential implies diode behavior of the depinning current. Applied currents to the right (left) tilt the vortex potential in opposite directions. Depinning occurs once the tilted vortex potential no longer exhibits minima, i.e., when the tilt becomes equal to the minimal (maximal) derivative of the potential in the absence of a bias current.


\subsection{Langevin equation for vortex motion and diode effect}

Following Ref.\ \cite{Rzchowski1990}, the dynamics of the vortex in the array can be described by a Langevin equation, which takes the same form as the Langevin equation for the phase difference of a single junction \cite{AmbegaokarHalperin1969}. Interestingly, this also allows for relating the diode effect of the array to a diode effect in single junctions \cite{Steiner2023}.

We focus on the vortex position $y$ along the $y$-axis, assuming overdamped Josephson junctions. In addition to the force $-\boldsymbol{\nabla} U$ due to the vortex potential, the vortex experiences a Lorentz force \cite{fazio2001quantum}
\begin{equation}
    \mathbf{F}_L = \gamma \Phi_0 (\mathbf{I}\times \hat{\mathbf{z}})/a 
\end{equation}
exerted by the externally applied current $I$ per lattice site. Here, $\Phi_0=h/2e$ is the superconducting flux quantum, $a$ the lattice constant of the Josephson junction array, and $\gamma$ a geometrical factor discussed below. Moreover, a friction force due to quasiparticle currents (friction constant $\alpha$), and the associated fluctuating Langevin force $\mathbf{f}$ act on the vortex. The Langevin equation for $y(t)$ takes the form 
\begin{equation}
    -\partial_y U - \alpha \dot y - \gamma \Phi_0 I/a  = f_y.
    \label{eq:langevin}
\end{equation}
The Langevin force is characterized by a zero average and correlator $\langle f_y(t)f_y(t')\rangle = 2\alpha k_BT \delta(t-t')$ at temperature $T$. For a large uniform array with a (small) average vortex density $n$, the electric field $E$ in the $x$-direction takes the form \cite{tinkham2004introduction}
\begin{equation}
    E = - \Phi_0 n  \bar{v},
    \label{eq: electric field}
\end{equation}
related to the the average drift velocity $\overline{v}$ of the vortex along $y$.


We now discuss the geometric factor $\gamma$. The current carried by the horizontal junctions exerts a Lorentz force on the vortices, which points along the $y$-direction, whereas the currents flowing through the diagonal junctions exert a force, which is rotated by $\pm 45^\circ$. While their $x$-components cancel (allowing us to focus on the $y$-direction), their $y$-components contribute to the total Lorentz force $F_L$. Computing the current distribution, we find $F_L=-\gamma \Phi_0 I/a$ with 
\begin{equation} 
    \gamma = \frac{E_J + \sqrt{2}E_D}{E_J+2E_D}
\end{equation}
in the limit of small phase offsets.

We can rewrite the Langevin equation in terms of a phase parameter defined through $\theta = 2\pi y/a$. In this case, the Langevin equation for the vortex maps directly on the Langevin equation for the phase across a single current-biased junction within the resistively shunted Josephson-junction  model,
\begin{equation}
    - \frac{2e}{\hbar}\partial_\theta U(\theta) - \frac{\hbar }{ 2e R} \dot\theta - I_\mathrm{bias} = \delta i
\end{equation}
with $\langle\delta i(t)\delta i(t')\rangle = (2k_BT/R) \delta(t-t')$. The mapping uses the identifications $R \leftrightarrow \Phi_0^2 / a^2\alpha$ for the shunt resistance and $I_\mathrm{bias} \leftrightarrow \gamma I$ for the bias current. The Josephson energy $U(\theta)$ can be directly identified with the vortex potential. Finally, the voltage $V=(\hbar/2e) \dot\theta $ across the single junction is related to the electric field $E$ in the uniform array according to $V \leftrightarrow E/(na)$. According to this mapping, the Josephson diode effect is encoded in the structure of $U(\theta)$ for both, single junctions and arrays. 


For numerical calculations, we discretize the time dependence of the vortex position $y$ and rewrite the Langevin equation in Eq.\ \ref{eq:langevin} in dimensionless units. At temperature $T=0$, it takes the form 
\begin{equation}
    -\frac{1}{2\pi }\frac{\dd \tilde U}{\dd \tilde{y}} - \frac{\dd \tilde{y}}{\dd \tilde{t}} - \gamma {\tilde I} = 0.
\end{equation}
Here we introduced  dimensionless variables (indicated by a tilde) through $U=E_J \tilde U$, $y = a \tilde{y}$, $t = (\alpha a^2/2\pi E_J) \tilde{t}$ as well as $I = I_0 \tilde I$, where $I_0 = 2\pi E_J/\Phi_0$ is the critical current per plaquette. We further use Eq.\ \ref{eq: electric field} to define a dimensionless voltage $\tilde{V}$ per plaquette through 
\begin{equation}
    V = \frac{2\pi E_J \Phi_0 n}{\alpha} \tilde{V}.
\end{equation}
The voltage $\tilde V$ can be obtained from the Langevin equation via
\begin{equation}
    \tilde V = -\overline{\frac{\dd \tilde{y}}{\dd \tilde{t}}},
\end{equation}
where the bar indicates a time average. The time-averaged vortex velocity in dimensionless units is obtained from the Langevin equation as a function of the bias current $\tilde I$ and two dimensionless parameters, the ratio $E_D/E_J$ of Josephson energies and the phase offset $\varphi_0$.

The calculated $V-I$ characteristic is displayed in Fig.\ \ref{fig: potential,vortex}(e).  For $E_D=0$ and $\Delta\varphi_0=0$, we find a current-voltage characteristic akin to that of a conventional single junction. Setting $E_D=E_J/2$ and $\Delta\varphi_0 = \pi/2$, the positive and negative depinning currents increase asymmetrically due to the increases in the maximum and minimum of $\partial_y U$, consistent with the diode effect observed in the experiment. 

\subsection{Depinning currents beyond the arctan-approximation} 

\begin{figure*}[t!]
\hspace*{-0.8cm}
    \includegraphics[scale = 0.65]{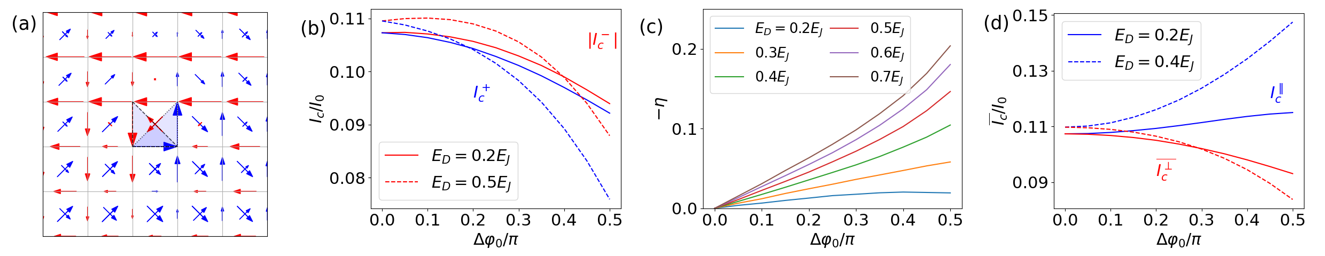}
    \caption{Depinning currents with phase relaxation. (a) Current distribution for a relaxed vortex configuration. The length of arrows  indicates the magnitude of the currents. Notice that in the central plaquette, the currents circulate around the highlighted triangles, reflecting the shift of the vortex minimum into the lower half of the plaquette. (b) Positive and negative depinning currents as a function of $\Delta \varphi_0$ for two values of $E_D$. (c) Diode efficiency defined in Eq.\ \ref{eq: diode efficiency} as a function of $\Delta \varphi_0$ for various values of $E_D$. (d) Comparison of depinning current parallel (in blue) and perpendicular (in red; averaged over current directions) to the applied in-plane magnetic field as a function of $E_D$ and $\Delta \varphi_0$. Parameters: $n_{max} = 10^5$,$\tau = 0.2$,$k=1200$. For horizontal currents: $I_{min} = 0.06I_0$, $I_{max} =0.12I_0$. For vertical currents: $I_{min} = 0.09I_0$, $I_{max} = 0.15I_0$.}
    \label{fig: critical current}
\end{figure*}

We now go beyond the $\arctan$ approximation of the vortex configuration (Eq.\ \ref{eq: vortex phase configuration}) and allow for relaxation  of the phase configuration, using a gradient-descent scheme. We tilt the washboard potential of the Hamiltonian in Eq.\ \ref{eq: Hamiltonian}, 
\begin{equation}
    H_I = H + \frac{\hbar}{2e} I \sum_{j=1}^N (\varphi_{1,j}-\varphi_{N,j})
\end{equation}
for an applied current $I$ in the $x$-direction. For small $I$, we minimize the energy for phase configurations containing a single vortex. We start with the phase configuration $ \varphi_{i,j}(x_0,y_0)$ in Eq.\ \ref{eq: vortex phase configuration} with an (unrelaxed) vortex located at  the center $(x_0,y_0)$ of the array, and gradually evolve the system towards the local minimum via a standard gradient-descent scheme. 

The system  monotonically converges to a stable solution, as long as the current $I$ remains below the depinning current $I_c$. Beyond the depinning current, the vortex keeps hopping along the $y$-direction. We start with a small current.
If the vortex remains in its initial plaquette after $n_{max}$ 'time' steps of length $\tau$, the current is below the depinning current. 
We then repeat the descent scheme for a slightly larger current, continuing in $k$ small increments up to a current density of $I_{\text{max}}$. We identify the depinning current $I_c$ with the current, for which the vortex no longer remains in the initial plaquette. 
We implement the procedure for currents of both signs.

Our results are shown in Fig.\ \ref{fig: critical current}. Fig.\ \ref{fig: critical current}(a) displays the current configuration for a relaxed vortex in the presence of  diagonal couplings and phase offset. The plaquette containing the vortex hosts diagonal currents. 
We observe that the current circulates only around the two lower triangles of the plaquette (as highlighted in the figure). This  indicates that the position of the vortex is shifted away from the center of the plaquette in the negative $y$-direction. This is consistent with the vortex potential shown in Fig.\ \ref{fig: potential,vortex}(b).

Figure \ref{fig: critical current}(b) shows the depinning currents $I_c^+$ and $I_c^-$ as a function of the phase offsets $\Delta\varphi_0$ for two values of the diagonal couplings $E_D$. The direction of the currents is indicated by color. We find that the depinning currents for both directions are reduced with increasing $\Delta\varphi_0$, but their asymmetry becomes stronger. The asymmetry is quantified by the diode efficiency
\begin{equation}
    \eta = 2 \frac{I_c^+ - I_c^-}{I_c^+ + I_c^-},
    \label{eq: diode efficiency}
\end{equation}
which is shown in Fig.\ \ref{fig: critical current}(c). The diode efficiency monotonically increases with both $E_D$ and $\Delta\varphi_0$. 

We also computed the depinning currents $I_c^{\parallel}$ in the $y$-direction, which do not exhibit a diode effect, consistent with the absence of a symmetry breaking in this direction. Moreover,    $I_c^{\parallel}$ increases with phase offset $\Delta\varphi_0$. This is in contrast to the depinning current $\overline{I_c^{\perp}}$ in the $x$-direction (averaged over current directions), which decreases with $\Delta\varphi_0$. This is shown in Fig.\ \ref{fig: critical current}(d). We note that unlike the arctan approximation (via the geometric factor $\gamma$), 
the phase-relaxation approach is no longer limited to small phase offsets. 

\subsection{Vortex-Antivortex symmetry of the diode efficiency: a peculiar characteristics of magnetochiral ratchets}
The observed diode efficiency is symmetric with respect to the out-of-plane magnetic field, i.e., $\eta(B_z) = \eta(-B_z)$.

This unprecedented observation marks an important difference between ordinary asymmetric pinning potentials (which can be obtained by engineering geometrically an asymmetry of the pinning potential) and the magnetochiral ratchet we present in this work.

\begin{figure*}[t!]
\centering
    \includegraphics[width=0.7 \textwidth]{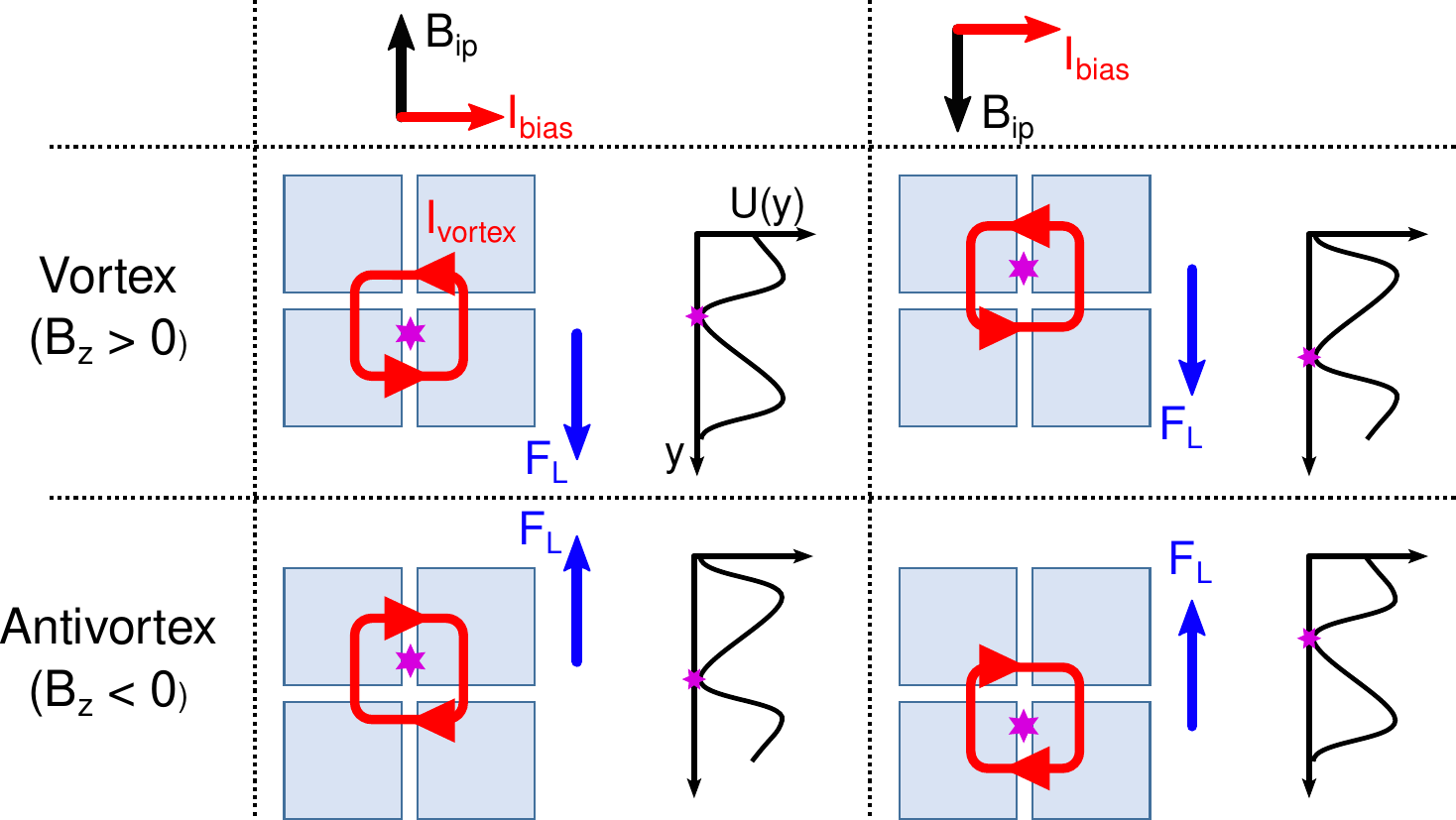}
    \caption{Vortex vs antivortex depinning.
    Qualitative shape of the currents and Lorenz forces for vortices (upper row) vs. antivortices (lower row). The columns differ by the polarity of the in-plane field ($\theta = -90^\circ$ in the left column vs. $\theta = 90^\circ$ in the right column), while the current bias direction is the same.
    For a given polarity of $B_\text{ip}$ the sign of rectification efficieny is the same for vortices and antivortices, while changing the polarity of $B_\text{ip}$ flips the sign of the rectification efficiency. The left column corresponds to the situation of lower critical current (for $\vec{B}_\text{ip}\times\vec{I}$ inside the page, the Lorentz force $\vec{F}_L$ pushes in the direction of the lower slope for both vortex and antivortex), the right column that for higher critical current.
  }
    \label{fig:Supp-vortex-antivortex-symmetries}
\end{figure*}

As discussed in the main text and in the previous section of this Supplement, the ratchet character of the pinning potential is due to spin-orbit terms in the Hamiltonian. This means that by inversion of the current density (i.e., changing a vortex with an antivortex) terms, the $\vec{B}_\text{ip}\times \vec{j}$ terms change sign, so that the magnetochiral tilt of the pinning potential changes sign, see sketch in Fig.~\ref{fig:Supp-vortex-antivortex-symmetries}.
This additional change of sign is the crucial point. In fact, for a given current bias polarity, also the Lorentz force changes sign. The two changes of sign are such that the depinning current is then \textit{the same} for vortices and antivortices. As a result, the rectification efficiency $\eta$ is expected to be symmetric in $B_z$, as we observe.

In stark contrast, for an ordinary ratchet-like pinning potential (namely, a potential engineered to be spatially asymmetric, without spin-orbit contribution) is the same for vortices and antivortices. This means that the opposite Lorentz force felt by vortices and antivortices leads to a different depinning current. The rectification is thus expected to be \textit{antisymmetric} in $B_z$.

The observation of a symmetric $\eta$ in Fig.~1 of the main text is therefore a new observation indicating a unique feature of magnetochiral ratchets.


\clearpage
\newpage

\section{Device layout and characterization}

\begin{figure*}[tb]
\centering
\includegraphics[width=\textwidth]{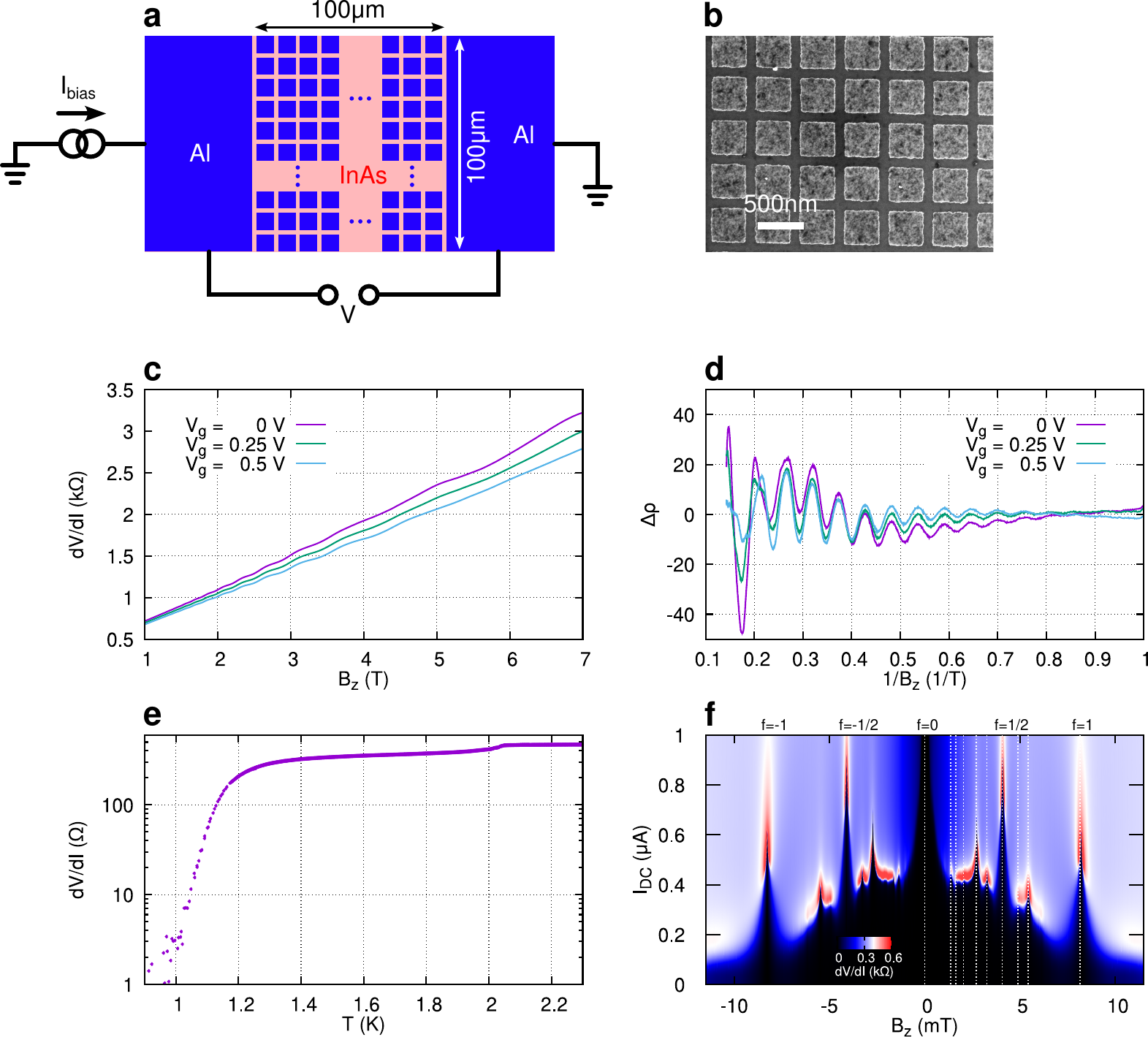}
\caption{\textbf{a,} Device geometry of devices A and B.
\textbf{b, } Scanning electron microscope image of device A
\textbf{c,} Two-point measurement of magnetoresistance in the normal state as a function of out-of-plane magnetic field. The temperature is $T\sim 200$~mK.
\textbf{d,} Magnetoresistance data after substraction of a quadratic polynomial background, plotted as a function of $1/B_z$. 
\textbf{e,} Temperature dependence of zero-bias differential resistance at zero magnetic field. 
\textbf{f,} Differential resistance at zero in-plane field and $T\sim 40$~mK measured as a function of out-of-plane field and dc bias current. Peaks of the depinning current are found at the indicated dotted lines at $f=B_z/B_0 = 0$, $1/6$, $1/5$, $1/4$, $1/3$, $2/5$, $2/3$, $1$. The measurement is performed with an ac-excitation of $10$~nA at frequency $777$~Hz.
}
\label{fig:Supp_device}
\end{figure*}

The used semiconductor/superconductor layer stack is described and characterized in the supplemental information of our previous work \cite{Reinhardt2024}, which also describes the details of fabrication. The layout of devices A and B is shown in Fig.~\ref{fig:Supp_device} \textbf{a}. The arrays consist of $200\times 200$ square islands with a lattice constant $a  = 500$~nm. The junctions between the islands have a length of $\sim 100$~nm. Measurements are performed in a dilution refrigerator with a base temperature of $T\sim 35$~mK. Details of the measurement setup can be found in the supplemental information of \cite{Reinhardt2024}. 
Unless stated otherwise, measurements are performed on device A. 

The electron density in the semiconductor quantum well can be found from magnetoresistance measurements performed with the array sample. The used magnetic fields are far above the critical field of the aluminum superconductor ($B_c \approx 100$~mT).
Fig.~\ref{fig:Supp_device} \textbf{c} shows the differential resistance of the array as a function of out-of-plane magnetic field for different gate voltages. Fig.~\ref{fig:Supp_device} \textbf{d} shows the same data after removing a second-order polynomial background with clearly visible Shubnikov de-Hass (SdH) oscillations. The density corresponding to the oscillations is $9.0\times 10^{11}$cm$^{-2}$ with no apparent dependence on gate voltage. This indicates that the SdHs probe the electron density in the semiconductor covered by Al, where the effect of the gate is screened. The obtained density is in excellent agreement with the value $n = 8.5\times 10^{11}$cm$^{-2}$ found by recent cyclotron resonance measurements on a similar wafer \cite{Chauhan2022}. The linear part of the magnetoresistance is caused by the two-terminal measurement, where voltage is probed on the terminals which are used to source the current. The two-terminal resistance will probe both the longitudinal resistance and the Hall resistance \cite{RikkenPRB1988}.

The temperature dependence of zero-bias differential resistance of the array is shown in Fig.~\ref{fig:Supp_device} \textbf{e}.
A small decrease of resistivity is found below the critical temperature of the film $T_{c,Al} \sim 2.1$~K. The onset of Josephson coupling is found below $1.1$~K.
The differential resistance at $T\approx 40$~mK as a function of out-of-plane field and bias current is shown in Fig.~\ref{fig:Supp_device} \textbf{b}. We find pronounced peaks of the depinning currents at the integer frustration $f:= B_z/B_0 =\pm 1$ with the matching field $B_0 = \pm \Phi_0 / a^2 \sim  8.2$~mT. Additional peaks are found at  various fractions of the frustration parameter, with the most pronounced peaks at $f=1/2$, $f=1/3$, and $f=2/3$.

The range of the vortex-vortex interaction in a 2D JJA is related to the magnetic penetration depth
\begin{equation}
\lambda_\perp = \frac{\Phi_0}{2\pi\mu_0 I_c}
    \label{penetration_depth}
\end{equation}
where $I_c$ is the critical current of a single junction. For a JJ of width $500$~nm we expect $I_c \sim 500$~nA in zero magnetic field, which yields $\lambda_\perp \sim 500$~\textmu m. As this largely exceeds the dimensions of the device, we do not have to take screening effects into account.
\clearpage \newpage
\section{IV-characteristics at low magnetic field}

\begin{figure*}[htb]
\centering
\includegraphics[width=\textwidth]{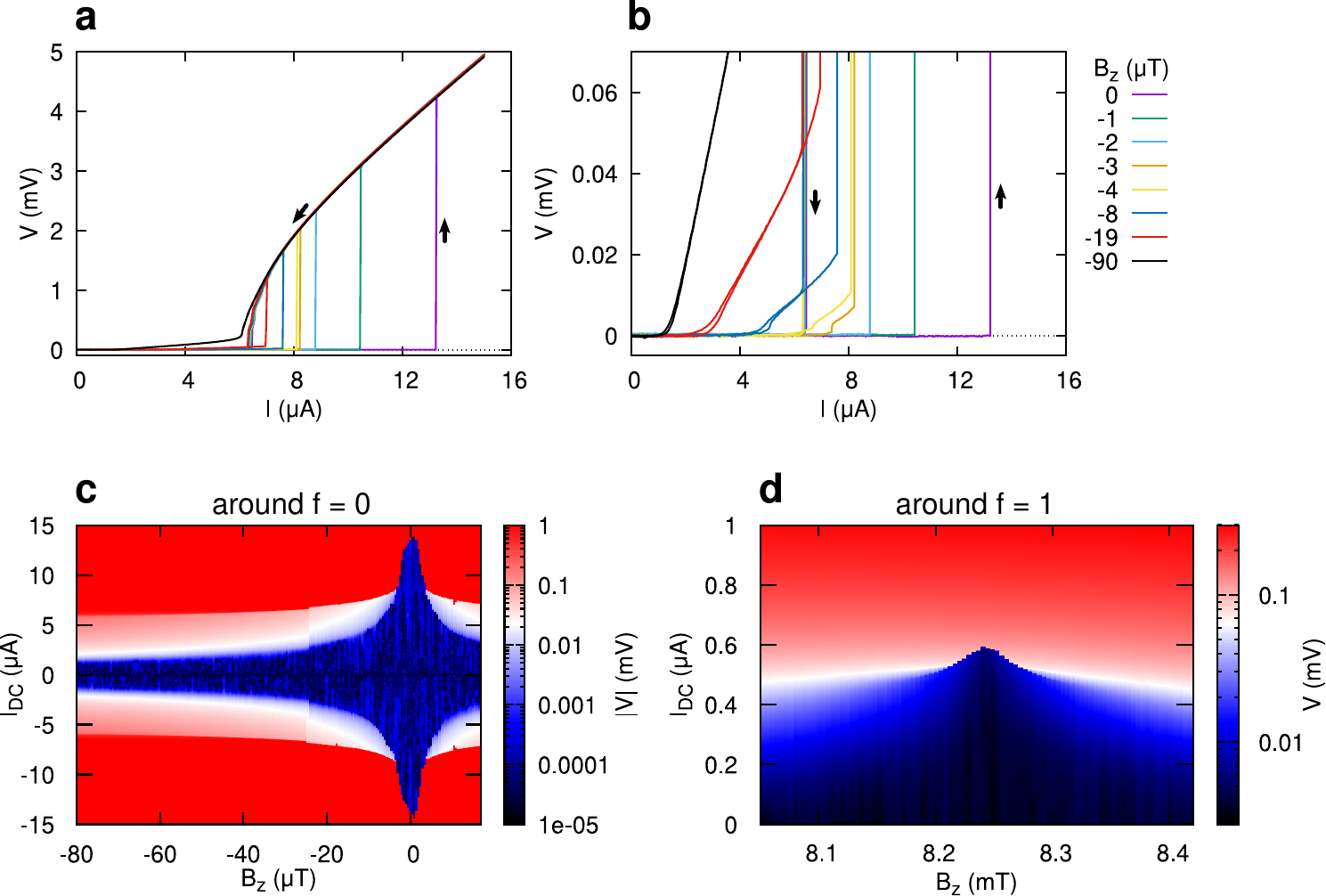}
\caption{\textbf{Current-voltage characteristics at zero in-plane magnetic field around $f=0$ and $f=1$}
\textbf{a,} $V(I)$ curves for low out-of-plane fields. Black arrows indicate the sweep direction, showing the hysteresis between up and down sweep of the current. Field-cooling is performed for every value of $B_z$ and $V(I)$ measurements always start at zero bias current.
\textbf{b,} Zoom into the low voltage part of \textbf{a}.
\textbf{c,} Color map of $\log(V(B_z, I))$ around zero frustration.
\textbf{d,} Color map of $\log(V(B_z, I))$ around integer frustration $f=1$.
}
\label{fig:Supp_low_field}
\end{figure*}

Fig.~\ref{fig:Supp_low_field} shows current-voltage characteristics obtained around zero and integer frustration. In order to avoid hysteresis from non-equilibrium flux pinning in the array and the superconducting contacts, we perform field cooling for each value of the out-of-plane field. 

The $V(I)$ curves feature discontinuous jumps with a large hysteresis between up and down sweep of the current. The hysteresis is likely caused by heating due do the the large power dissipation in the resistive state.  For fields below $\sim 2$~$\mu$T we observe a direct jump between the zero voltage state (black/dark blue in Fig.~\ref{fig:Supp_low_field} \textbf{c}) and the voltage state. For higher out-of-plane field we find a regime of flux creep with approximately constant slope of $V(I)$, followed by the jump into the resistive state. The discontinuous jump into the resistive state is also found around integer frustration, as shown in Fig.~\ref{fig:Supp_low_field} \textbf{d}.
As expected for vortex depinning, the observed depinning current densities are a factor of 10 below the  Josephson critical current densities obtained using single junction devices made with material from the same wafer \cite{Reinhardt2024}. 
We conclude that the measured critical current is always caused by depinning of vortices, even in nominally zero out-of-plane field.
The field $\Phi_0/A \approx 200$~nT corresponding to a single vortex in the array is below the resolution of the out-of-plane field used in our measurements.

    

\clearpage \newpage

\section{Additional data on non-reciprocal depinning current}

\begin{figure*}[htb]
    \centering
    \includegraphics[width=\textwidth]{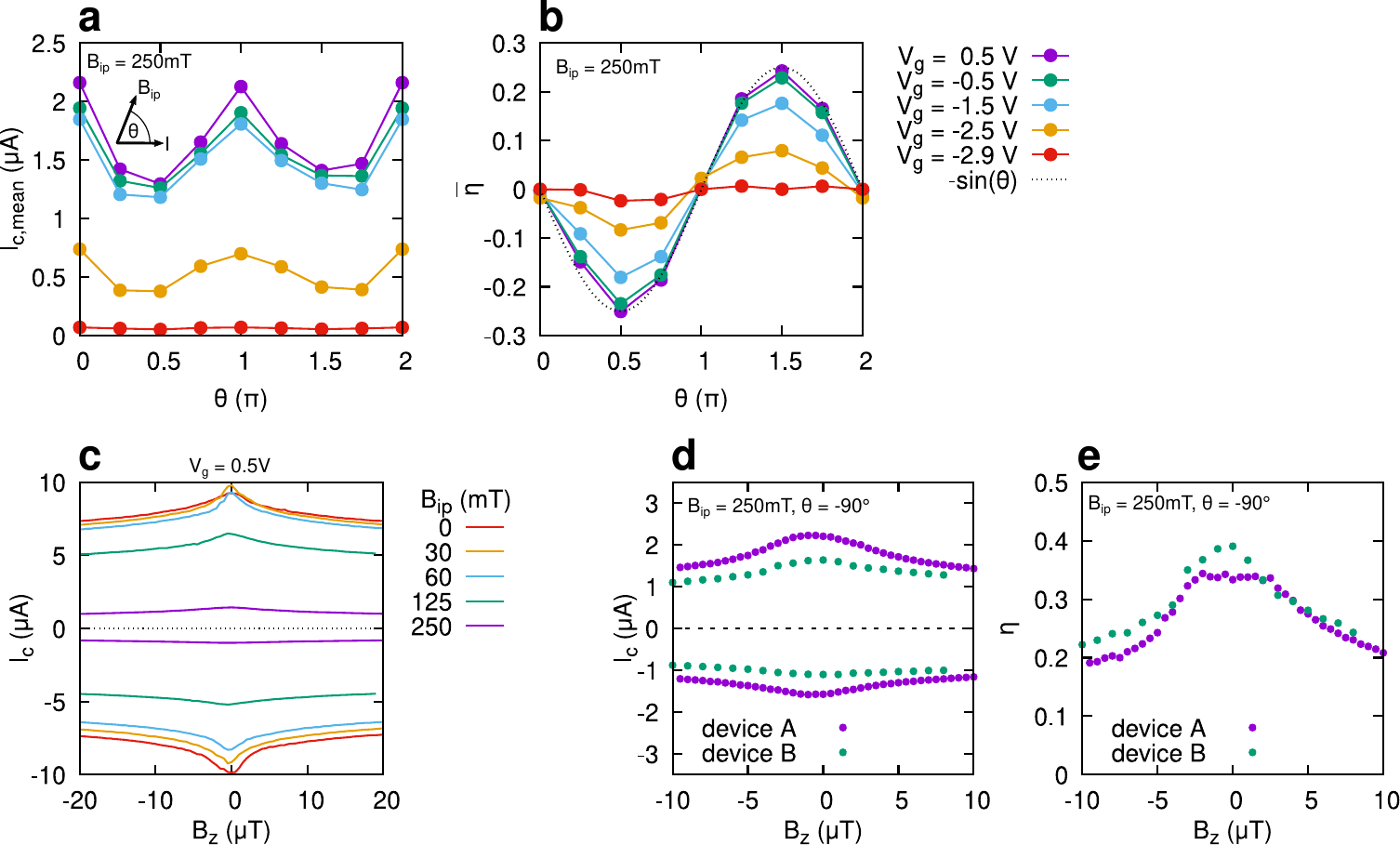}
    \caption{\textbf{Additional data on non-reciprocal depinning current.}
    \textbf{a,} Angle dependence of the mean depinning current $1/2(I_c^+ + I_c^-)$. The depinning current is averaged over the range $|B_z| < 20$~\textmu T.
    \textbf{b,} Angle dependence of the averaged diode efficiency $\bar{\eta}$.
    \textbf{c,} Depinning current as a function of out-of-plane field for different magnitudes of the in-plane field ($\theta = -90$~$^\circ$).
    \textbf{d,e,} Comparison of depinning current and diode efficiency for the nominally identical devices A and B.
    }
    \label{fig:Fig_Supp_diode_extra_data}
\end{figure*}

\begin{figure*}[htb]
    \centering
    \includegraphics[width=\textwidth]{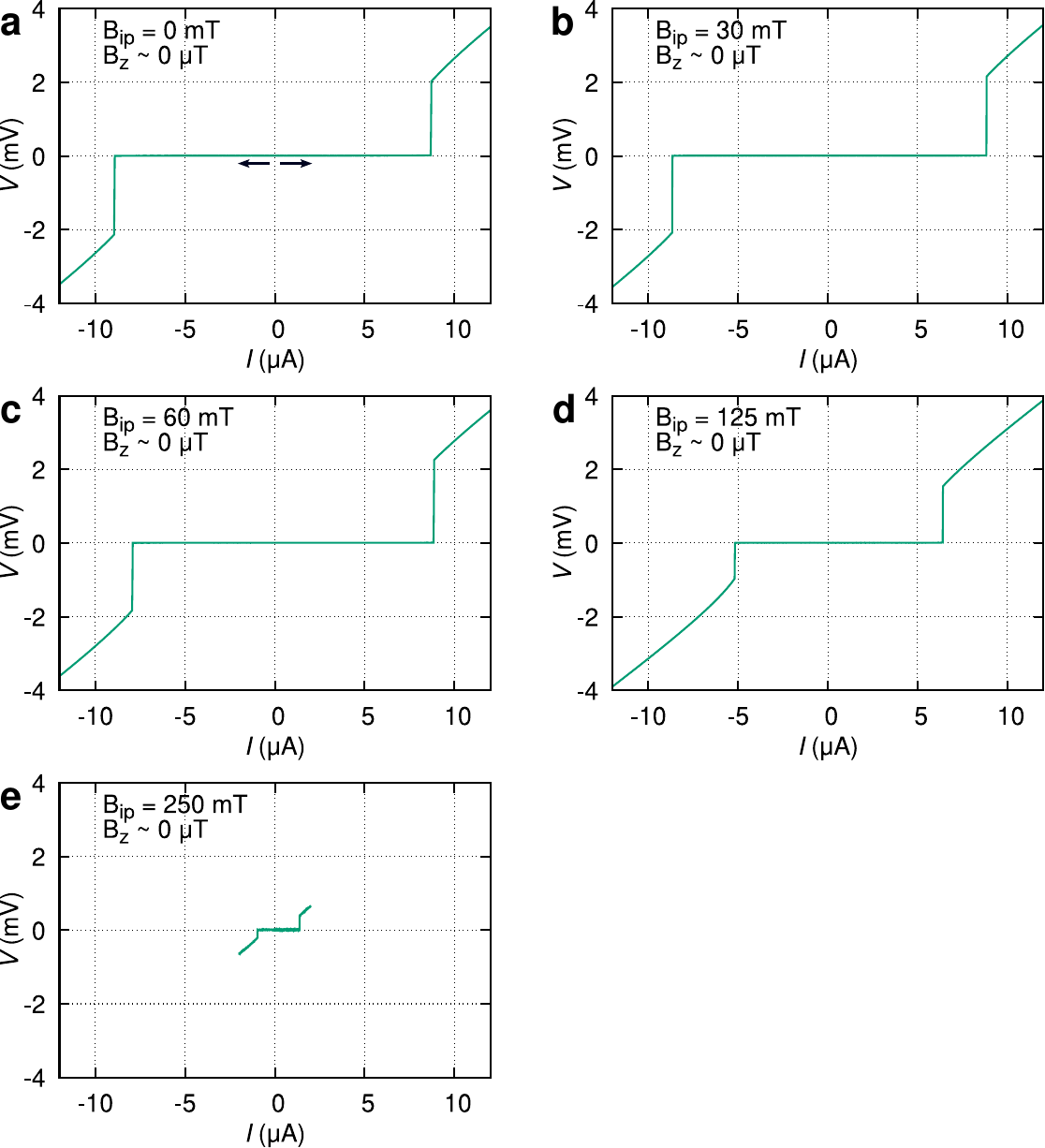}
    \caption{\textbf{Example $I(V)$ curves for different values of $B_\text{ip}$.}
    This data corresponds to data shown in Fig.~1\textbf{e} of the main text.
    For all measurements the in-plane field is applied with an angle $\theta = -90^\circ$ and the out-of-plane field is close to zero. The gate voltage is set to $V_g = 0.5$~V. For both polarities of the bias current the sweep always starts at zero, as indicated with black arrows in panel \textbf{a}.
    The applied in-plane fields are $0$~mT for \textbf{a}, $30$~mT for \textbf{b}, $60$~mT for \textbf{c}, $125$~mT for \textbf{d}, and $250$~mT for \textbf{f}.
    }
    \label{fig:Supp-raw-IVs}
\end{figure*}

Fig.~\ref{fig:Fig_Supp_diode_extra_data} \textbf{a,b} show the angle dependence of the depinning current and the diode efficiency for an in-plane field of $250$~mT. 
The averaged depinning current shows pronounced maxima when the field is parallel to the current ($\theta = 0,\pi$) and minima when the field is perpendicular to the current ($\theta = \pi/2, 3\pi/2$).
The angle dependence of the averaged diode efficiency shows the sinusoidal behaviour $\bar{\eta} \sim -\sin(\theta)$.
Fig.~\ref{fig:Fig_Supp_diode_extra_data} \textbf{c} shows the depinning current as as a function of the in-plane magnetic field, corresponding to Fig.~1 \textbf{e} of the main text.
We compare the non-reciprocal depinning current of devices A and B at an in-plane field of $B_\text{ip} = 250$~mT (Fig.~\ref{fig:Fig_Supp_diode_extra_data} \textbf{d,e}).
The depinning current is slightly lower for device B. The magnitude of the diode efficiency of the both devices is nearly equal.

Fig.~\ref{fig:Supp-raw-IVs} shows raw $V(I)$ data corresponding to the data of Figs.~1 \textbf{a,e} of the main text and Fig.~\ref{fig:Fig_Supp_diode_extra_data} \textbf{c}. The shown $V(I)$ curves correspond to zero out-of-plane field. The difference of critical current for positive and negative polarity becomes clearly visible for $B_\text{ip} > 60$~mT.

\clearpage \newpage
\section{Dependence of the depinning current and diode efficiency on the field sweep direction}
\begin{figure*}[htb]
    \centering
    \includegraphics[width=\textwidth]{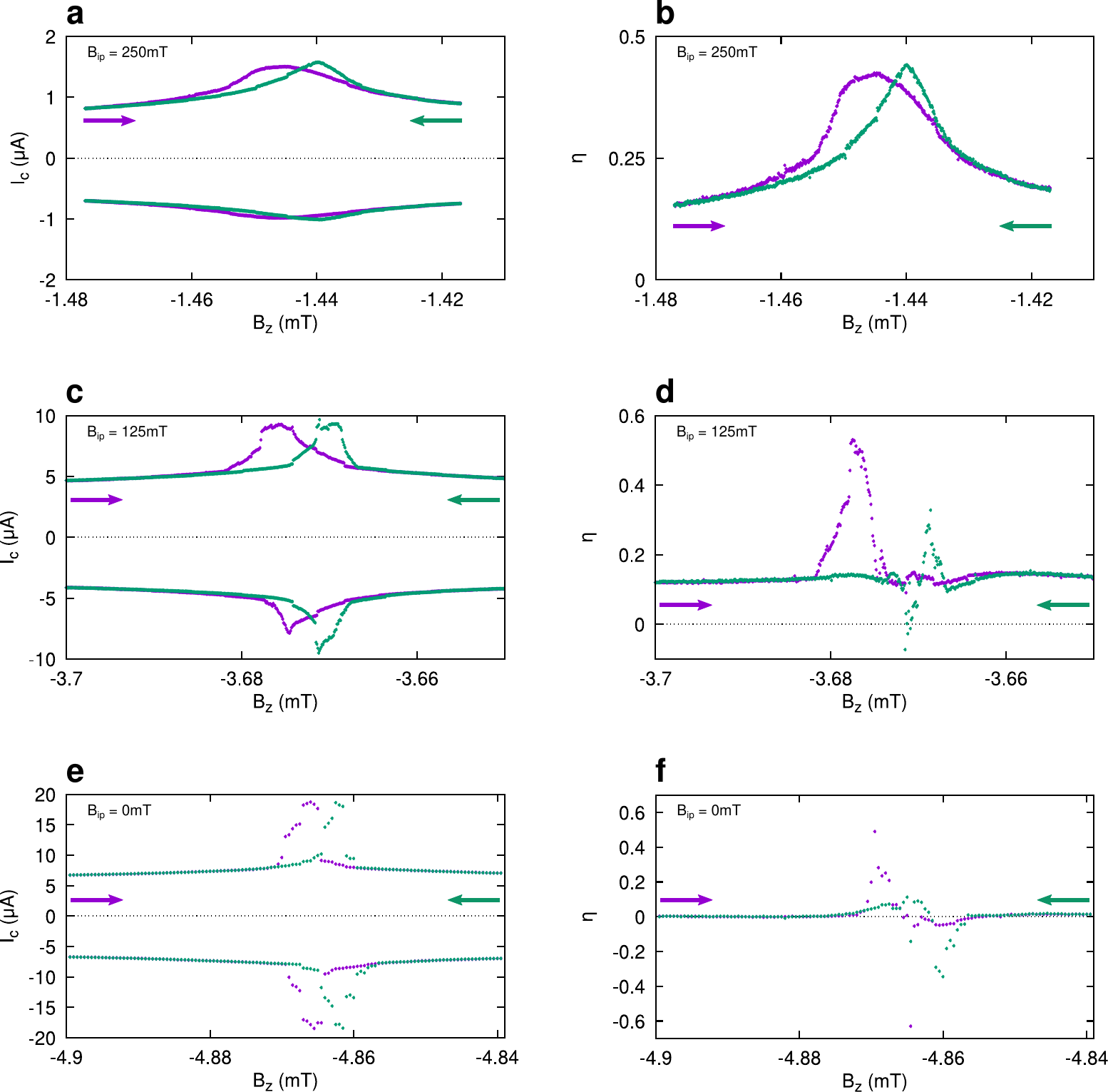}
    \caption{\textbf{Dependence of depinning current and diode efficiency on the sweep direction of the out-of-plane magnetic field.}
    Out-of-plane field dependence of the depinning current $I_c$ (left panels \textbf{a,c,e}) and diode efficiency (right panels \textbf{b,d,f}) for $B_\text{ip}=250$~mT (top, \textbf{a,b}), $B_\text{ip}125$~mT (middle,  \textbf{c,d}), $B_\text{ip}=0$~mT (bottom, \textbf{e,f}). 
    The sweep direction of the out-of-plane field is indicated with purple/green arrows.
    In all measurements the in-plane magnetic field is applied in negative $y$-direction ($\theta = -\pi/2$) and the temperature is below $50$\,mK.
    No offsets are subtracted from $B_z$, in contrast to the graphs in the main text. The offset fields are caused by misalignment of the in-plane fields and flux-pinning in the superconducting magnet coils.
    }
    \label{fig:Fig_Supp_diode_hysteresis}
\end{figure*}

Changing the out-of-plane magnetic field $B_z$ while keeping $T \ll T_c$ will result in a non-equilibrium configuration of vortices. When sweeping the out-of-plane field around zero, we expect that the depinning currents will display an hysteretic dependence on the out-of-plane magnetic field.
Fig.~\ref{fig:Fig_Supp_diode_hysteresis} shows measurements of $I_c$, for both positive and negative sweep direction of $B_z$, performed around $B_z = 0$~\textmu T. The shape of $\eta(B_z)$ clearly depends on the sweep direction for $B_z$ close to zero, while at larger magnitude of $B_z$, the critical currents and diode efficiency become independent of the sweep direction. 
Interestingly we observe an anti-hysteretic effect: sweeping $B_z$ from negative to positive values, one observes a peak for $B_z<0$ (and vice versa). We do not have so far an explanation for the puzzling inversion of the hysteretic effect. However, we stress that upon thermal annealing (i.e., after setting a new $B_z$ value, the temperature is raised above $T_c$ and then the sample is cooled down at constant $B_z$) the hysteresis disappears and the curves become symmetric. This is the case for example of Fig.1\textbf{e} of the main text.

These observations provide an explanation for the slightly different shape of $\eta(B_z)$ in the panels Fig.~1\textbf{e} and Fig.~2\textbf{a} of the main text. In Fig.~2\textbf{a} the data is recorded like the purple curves in Fig.~\ref{fig:Fig_Supp_diode_hysteresis} by sweeping $B_z$ in positive direction while the temperature is close to the base temperature of the cryostat (i.e., no thermal annealing procedure is applied). Thus, curves in Fig.~2\textbf{a} of the main text appear not centered around zero.
In contrast, in Fig.~1\textbf{e} an annealing procedure is applied  for every value of $B_z$, so that vortices form a low-energy equilibrium configuration. This procedure results in curves of $\eta(B_z)$ with larger symmetry around zero out-of-plane field and absence of hysteresis.


\clearpage\newpage
\section{The role of diagonal coupling for the vortex ratchet effect: A control experiment}

\begin{figure}[h]
  \centering
  \includegraphics[width=\textwidth]{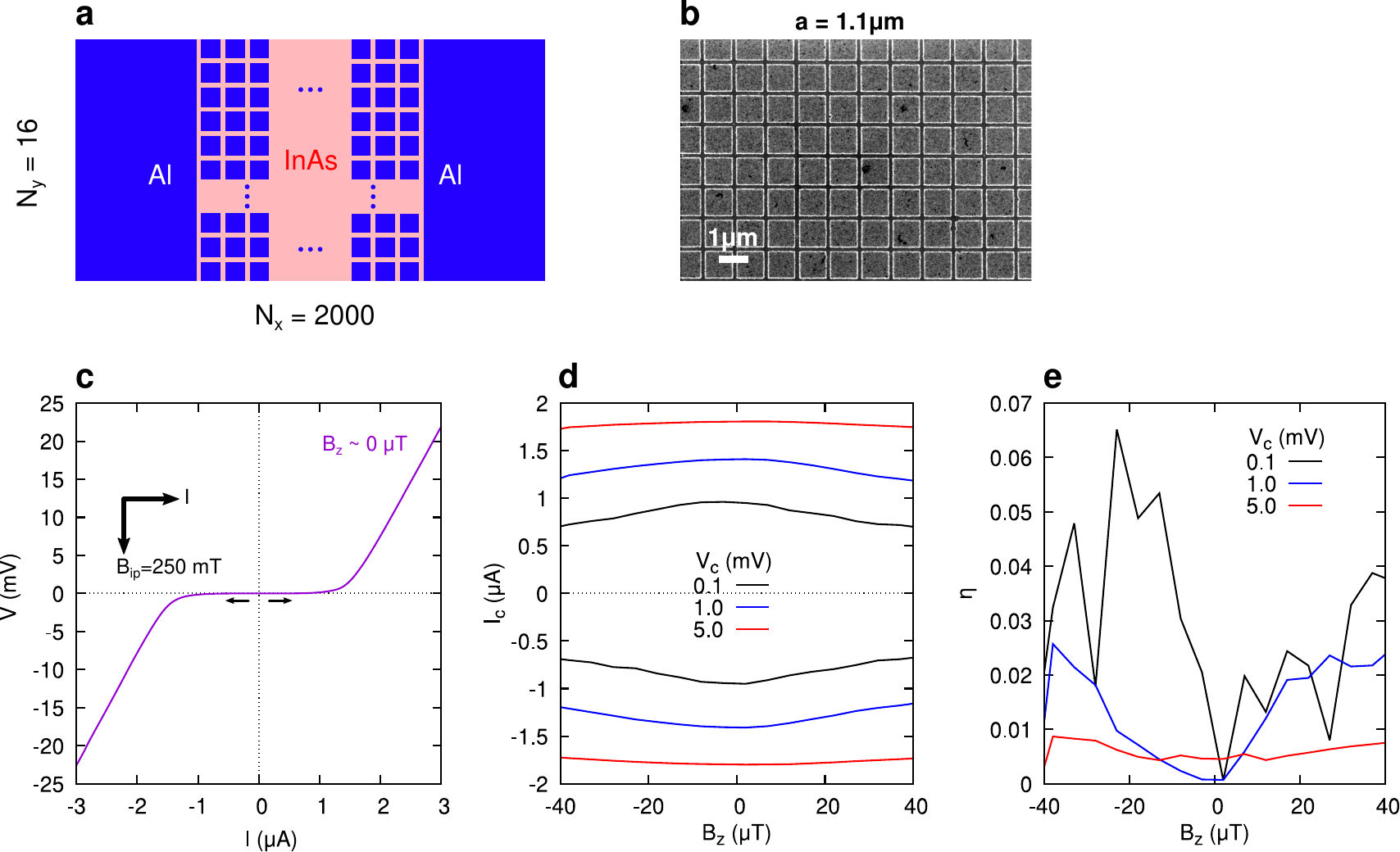}
  \caption{
    \textbf{a,} Geometry of The JJA device C.
    The Array consists of $16 \times 2000$ islands with two superconducting
    aluminum contacts. Transport experiments are performed using a quasi
    four-point configuration. Due to manufacturing reasons the array is actually formed by 20 shorter arrays of
    size $16 \times 100$. The individual arrays are connected by superconducting Al electrodes.
    \textbf{b,} SEM image of device C. The aluminum islands of size $1 \times
    1$~\textmu m$^2$ are separated by junctions of length $\sim 100$~nm.
    The lattice constant is $a = 1.1$\textmu m.
    \textbf{c,} $V(I)$ characteristic of device C obtained at $B_\text{ip} =
    250$~mT. The out-of-plane field is close to $0$~\textmu T and
    the gate voltage is set to $V_g = 0$~V.
    The sweep direction for positive and negative polarity of the bias current
    is indicated with black arrows.
    \textbf{d,} Critical current as a function of $B_z$ for different values of
    the voltage $V_c$ (see text).
    \textbf{e,} Rectification efficiency corresponding to the critical currents
    shown in panel \textbf{d}.
  }
  \label{fig:ratchet-control}
\end{figure}

In the used square array geometry the ratio of next-nearest to
nearest-neighbor coupling depends on the length scales of the Andreev bound
states  and the
array lattice constant. As the coherence length $\xi_0$ typically exceeds the
mean free path $l$, the spatial extend of ABS in our devices is ultimately
limited by the mean 
free path $l$ in the two-dimensional electron gas. A good metric to compare different square array geometries is then given by the
parameter $l / a$, i.e., the ratio between mean free path
 and
array lattice constant. The larger this ratio, the larger the next-nearest neighbor coupling in comparison with the nearest neighbor one.
In this section, we present a control experiment made using a new array device (device C) with strongly reduced
diagonal coupling, obtained increasing the Al island size and thus $a$. The device C geometry is presented in Fig.~\ref{fig:ratchet-control}\textbf{a,b}.
This JJA device is fabricated using the Al/InAs heterostructure M10-08-18.2, which was characterized in our previous work \cite{baumgartner2020}. The maximum mean free path in this structure is
$l_\text{max}\sim 270$~nm, observed at an electron density of $n = 0.5 \times 10^{12}$~cm$^{-2}$. The array with a shape of $16 \times
2000$ islands has a lattice constant of $a = 1.1$\textmu m.
For device C the ratio $l_\text{max}/a \sim 0.25$ is much lower compared to devices A and B.
We perform $V(I)$ measurements with an in-plane magnetic field with magnitude
$B_\text{ip} = 250$~mT, which is applied perpendicular to the direction
of the bias current. Fig.~\ref{fig:ratchet-control}\textbf{c} shows a $V(I)$
measurement performed close to zero out-of-plane magnetic field at zero gate
voltage. For both polarities the bias current is swept starting at $I = 0$.
The curve does not show the pronounced step at the critical current, which is
visible in measurements of devices A and B. Yet, it is possible to define a
critical current, e.g., as the current where the voltage exceeds a certain critical value $V_c$.
The resulting $I_c(B_z)$ curves for different values of the critical voltage
$V_c$ are shown in Fig.~\ref{fig:ratchet-control}\textbf{d}. Similar to devices A
and B, a maximum of $I_c$ is found around zero out-of-plane field.
The corresponding rectification is shown in Fig.~\ref{fig:ratchet-control}\textbf{d}. For low values
of $V_c$, the value of $\eta$ shows large fluctuations, while at the larger value
$V_c = 5$~mV the rectification efficiency is
always below $1$~\%. Compared to devices A and B where $\eta \gtrsim
0.35$ for the same parameters this clearly shows that the ratchet effect is present, but
strongly suppressed in device C. This provides a strong empirical evidence for the role of the diagonal coupling in the mechanism of the magnetochiral vortex ratchet effect.

\clearpage \newpage
\section{Additional data on non-reciprocal resistance}
\begin{figure*}[htb]
\centering
\includegraphics[width=\textwidth]{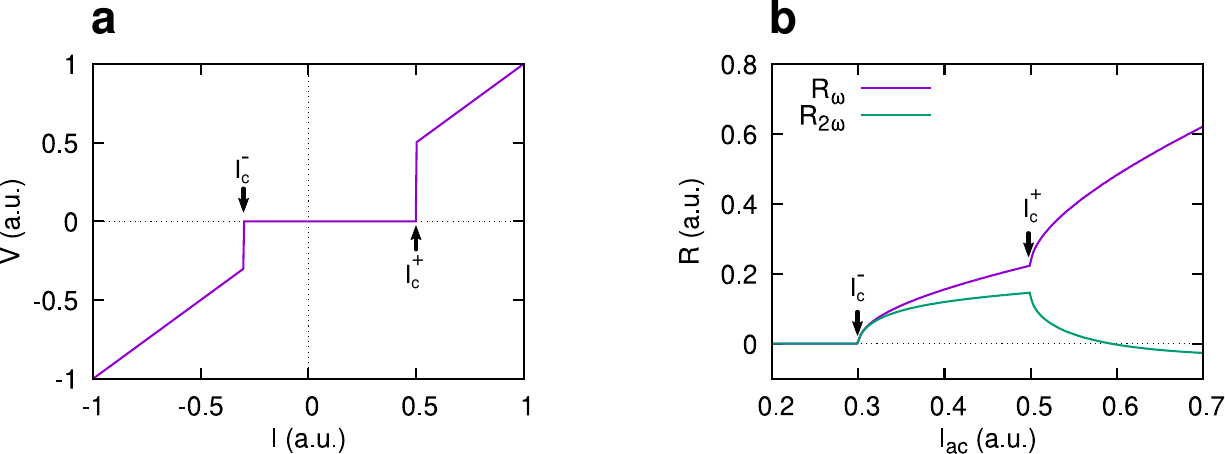}
\caption{\textbf{Simulation of non-reciprocal resistance.} 
\textbf{a}, Non-reciprocal $I(V)$ characteristic with $\eta = \Delta I_\text{c} / I_\text{c,mean} = 0.5$.
\textbf{b}, Numerically calculated first and second harmonic of resistance for an ac current bias with amplitude $I_\text{ac}$.
}
\label{fig:Fig_Supp_2omega_simulation}
\end{figure*}

Fig.~\ref{fig:Fig_Supp_2omega_simulation} \textbf{b} shows the expected behavior of $R_\omega$ and $R_{2\omega}$ for the $V(I)$ curve shown in Fig.~\ref{fig:Fig_Supp_2omega_simulation} \textbf{a}  with $I_c^+  > |I_c^-|$.
$R_{2\omega}$ is positive when the amplitude $I_{ac}$ of the sinusoidal excitation current is in the rectification window between negative and positive depinning current.

\begin{figure*}[htb]
\centering
\includegraphics[width=\textwidth]{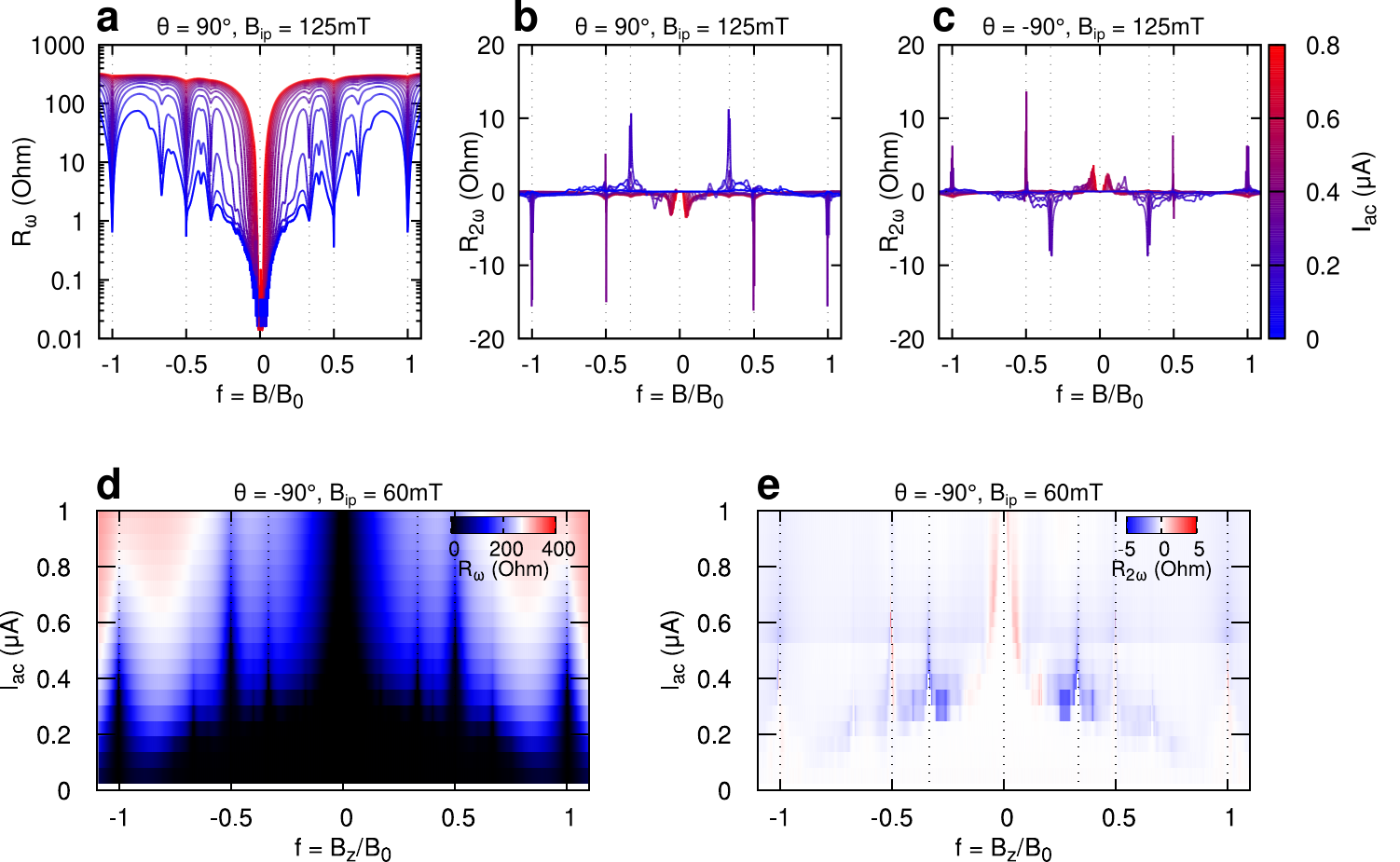}
\caption{\textbf{Additional measurements of non-reciprocal resistance as a function of frustration.} 
\textbf{a-c}, Same data as in Fig.~4 \textbf{a-c} of the main text showing $R_\omega(f)$ and $R_{2\omega}(f)$. Different colors correspond to different ac current excitations.
\textbf{d,e}, First and second harmonic of resistance measured at $B_\text{ip} = 60$~mT. $R_{2\omega}$ shows the sign-reversal in a region of frustration around $f=1/3$.
}
\label{fig:Fig_Supp_2omega_extra_data}
\end{figure*}

Fig.~\ref{fig:Fig_Supp_2omega_extra_data} \textbf{a-c} show linetraces of $R_\omega(f)$ and $R_{2\omega}(f)$, corresponding to Fig.~4 \textbf{a-c} of the main text.
No pronounced peak is present at $f=0$, as the maximum excitation current $I_{ac} = 0.8$~\textmu A is far below the depinning currents at $f=0$.
Additional measurements of $R_\omega$ and $R_{2\omega}$ are performed at $B_\text{ip} = 60$~mT. The resulting data is shown in Fig.~\ref{fig:Fig_Supp_2omega_extra_data} \textbf{d,e}. As for the case with $B_\text{ip} = 125$~mT, we find a pronounced sign-reversal of $R_{2\omega}$ around $f=1/3$.

\begin{figure*}[htb]
\centering
\includegraphics[width=0.5\textwidth]{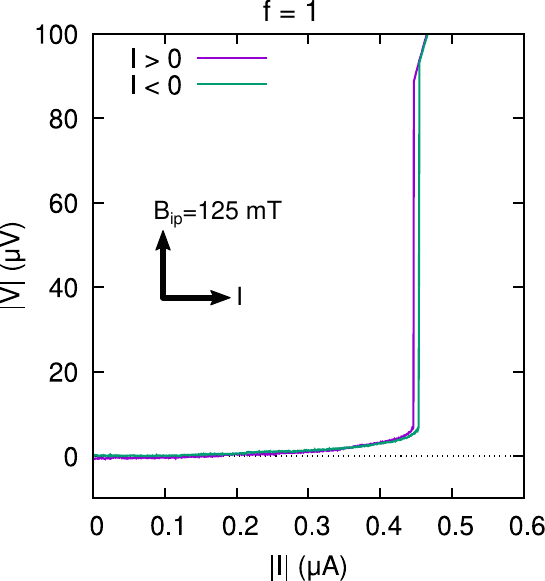}
\caption{\textbf{Non-reciprocal $V(I)$ at integer frustration $f=1$.} 
This $V(I)$ data corresponds to the $f=1$ data point of the data presented in Fig.~4\textbf{f} ($B_\text{ip} = 125$~mT, $\theta = 90^\circ$). The rectification efficiency is $\eta \sim -2$\%.
}
\label{fig:Supp-IV-f=1}
\end{figure*}

Fig.~\ref{fig:Supp-IV-f=1} shows the $V(I)$ curve corresponding to the $f=1$ data point presented in Fig.~4\textbf{f} of the main text ($B_\text{ip} = 125$~mT, $\theta = 90^\circ$).
While the rectification efficiency is much smaller in comparison to the $f\sim 0 $ case, a difference in critical currents for positive and negative polarity of the bias current is clearly visible.

\section{Data availability}
The data that support the ﬁndings of this study are available at the
online depository EPUB of the University of Regensburg, with the
identiﬁer DOI: 10.5283/epub.59682.

\end{document}